\documentclass[camera,10pt]{jpaper} 
\usepackage[noadjust]{cite}
\usepackage{amsmath,amssymb,amsfonts}
\usepackage{algorithmic}
\usepackage{graphicx}
\usepackage{textcomp}
\usepackage{stackengine}
\usepackage{hyperref}
\usepackage{balance}
\usepackage{url}
\def\BibTeX{{\rm B\kern-.05em{\sc i\kern-.025em b}\kern-.08em
    T\kern-.1667em\lower.7ex\hbox{E}\kern-.125emX}}

\hypersetup{%
    unicode=true,%
    pdfstartview={FitH},%
    colorlinks=true,%
    citecolor=black,%
    filecolor=black,%
    linkcolor=black,%
    urlcolor=black}

\usepackage{float}
\usepackage{xurl}
\usepackage{capt-of}
\usepackage{stfloats} 

\usepackage{atbegshi}
\usepackage{datetime2} 

\newif\ifuseversions
\useversionsfalse
\newcommand{\versionnum}[0]{4.0}
\ifuseversions
  \def\parsepdfdatetime#1:#2#3#4#5#6#7#8#9{%
    \def\theyear{#2#3#4#5}%
    \def\themonth{#6#7}%
    \def\theday{#8#9}%
    \parsepdftime
  }

  \def\parsepdftime#1#2#3#4#5#6#7\endparsepdfdatetime{%
    \def\thehour{#1#2}%
    \def\theminute{#3#4}%
    \def\thesecond{#5#6}%
    \ifstrequal{#7}{Z}
    {%
      \def\thetimezonehour{+00}%
      \def\thetimezoneminute{00}%
    }%
    {%
      \parsepdftimezone#7%
    }%
  }

  \def\parsepdftimezone#1'#2'{%
    \def\thetimezonehour{#1}%
    \def\thetimezoneminute{#2}%
  }

  \newcommand*{\thetimezone}{\thetimezonehour:\thetimezoneminute}
  \expandafter\parsepdfdatetime\pdfcreationdate\endparsepdfdatetime

  \newcommand{\version}[1]{\emph{Version #1 (Built:~\today~@ \DTMcurrenttime~UTC\thetimezone)}}
  \newcommand{\versionstring}{\textcolor{blue}{\normalsize\textit{\version{\versionnum}}}}

    \AtBeginShipout{%
    \AtBeginShipoutAddToBox{%
        \put(110,50){\versionstring}%
    }%
    }%
\fi

\usepackage{pifont}
\newcommand{\cmark}{\ding{51}}%
\newcommand*\circled[1]{\raisebox{.5pt}{\textcircled{\raisebox{-.9pt} {#1}}}}

\usepackage{array}
\newcolumntype{L}[1]{>{\raggedright\let\newline\\\arraybackslash\hspace{0pt}}m{#1}}
\newcolumntype{C}[1]{>{\centering\let\newline\\\arraybackslash\hspace{0pt}}m{#1}}
\newcolumntype{R}[1]{>{\raggedleft\let\newline\\\arraybackslash\hspace{0pt}}m{#1}}

\usepackage[resetlabels,labeled]{multibib}
\newcites{S}{Survey Sources} 

\usepackage[font={small,bf}]{caption}
\usepackage[shortlabels]{enumitem}
\setlist[itemize]{noitemsep,itemsep=0pt,parsep=0pt,topsep=0pt,partopsep=2pt,leftmargin=1em}
\setlist[enumerate]{noitemsep,itemsep=0pt,parsep=0pt,topsep=0pt,partopsep=2pt,leftmargin=1em}

\definecolor{grey}{rgb}{0.7, 0.6, 0.6}
\definecolor{darkgrey}{rgb}{0.5, 0.4, 0.4}
\definecolor{amber}{rgb}{1.0, 0.49, 0.0}
\definecolor{darkamber}{rgb}{0.9, 0.49, 0.0}
\definecolor{extradarkamber}{rgb}{0.7, 0.39, 0.0}
\definecolor{darkgreen}{rgb}{0.0, 0.2, 0.13}
\definecolor{darkbyzantium}{rgb}{0.36, 0.22, 0.33}
\definecolor{darkseagreen}{rgb}{0.56, 0.74, 0.56}
\definecolor{darkspringgreen}{rgb}{0.09, 0.45, 0.27}
\definecolor{dollarbill}{rgb}{0.52, 0.73, 0.4}
\definecolor{darkcerulean}{rgb}{0.03, 0.27, 0.49}

\newcommand{\xmt}[1]{#1}

\newcommand{\mpo}[1]{#1}
\newcommand{\mpt}[1]{#1}
\newcommand{\mph}[1]{#1}
\newcommand{\mpf}[1]{#1}
\newcommand{\mpg}[1]{#1}
\newcommand{\mpi}[1]{#1}
\newcommand{\mpj}[1]{#1}
\newcommand{\mpk}[1]{#1}
\newcommand{\mpl}[1]{#1}
\newcommand{\mpm}[1]{#1}
\newcommand{\mpn}[1]{#1}

\newcommand{\amb}[1]{#1}
\newcommand{\amc}[1]{#1}
\newcommand{\amd}[1]{#1}
\newcommand{\ame}[1]{#1}
\newcommand{\amf}[1]{#1}
\newcommand{\amg}[1]{#1}
\newcommand{\amh}[1]{#1}
\newcommand{\ami}[1]{#1}
    
\newcommand{\hcfirst}[0]{$\mathrm{HC}_{\mathrm{first}}$}
\newcommand{\hcfirstbold}[0]{$\mathrm{\textbf{HC}}_{\mathrm{\textbf{first}}}$}

\newcommand{\trcd}[0]{$t_{RCD}$}
\newcommand{\tras}[0]{$t_{RAS}$}
\newcommand{\trc}[0]{$t_{RC}$}
\newcommand{\trefi}[0]{$t_{REFI}$}
\newcommand{\trfc}[0]{$t_{RFC}$}

\usepackage{bm}

\newcommand{\btrefi}[0]{$\bm{t_{REFI}}$}
\newcommand{\btrfc}[0]{$\bm{t_{RFC}}$}

\begin{document}

\bstctlcite{IEEEexample:BSTcontrol}
\bstctlcite[@auxoutS]{IEEEexample:BSTcontrol}

\title{Rethinking the Producer-Consumer Relationship\\in Modern DRAM-Based Systems}

\newcommand{\ethz}{{\large$^1$}} 
\newcommand{\tud}{{\large$^2$}}
\author{ \vspace{-2ex}\\%
Minesh Patel\ethz{}\hspace{0.15in}%
Taha Shahroodi\tud{}$^,$\ethz{}\hspace{0.15in}%
Aditya Manglik\ethz{}\hspace{0.15in}%
A. Giray Ya{\u{g}}l{\i}k{\c{c}}{\i}\ethz{}\hspace{0.15in}\\%
Ataberk Olgun\ethz{}\hspace{0.15in}%
Haocong Luo\ethz{}\hspace{0.15in}%
Onur Mutlu\ethz{}\vspace{2mm}\\%
\textit{\ethz{}ETH Z{\"u}rich\hspace{0.12in}\tud{}TU Delft}%
\\}

\maketitle

\begin{abstract}
    Generational improvements to commodity DRAM throughout half a century have long
solidified its prevalence as main memory across the computing industry. However,
overcoming today's DRAM technology scaling challenges requires new solutions
driven by both DRAM producers and consumers. In this paper, we observe that the
separation of concerns between producers and consumers specified by
industry-wide DRAM standards is becoming a liability to progress in addressing
scaling-related concerns.

To understand the problem, we study four key directions for overcoming DRAM
scaling challenges using system-memory cooperation: (i) improving memory access
latencies; (ii) reducing DRAM refresh overheads; (iii) securely defending
against the RowHammer vulnerability; and (iv) addressing worsening memory
errors. We find that the single most important barrier to advancement in all
four cases is the consumer's lack of insight into DRAM reliability. Based on an
analysis of DRAM reliability testing, we recommend revising the separation of
concerns to incorporate limited information transparency between producers and
consumers. Finally, we propose adopting this revision in a two-step plan,
starting with immediate information release through crowdsourcing and
publication and culminating in widespread modifications to DRAM standards.
\end{abstract}

\newcommand\blfootnote[1]{%
  \begingroup
  \renewcommand\thefootnote{}\footnotetext{#1}%
  \addtocounter{footnote}{-1}%
  \endgroup
}

\section{Introduction}
\label{sec:intro}
\label{newsec:intro}

Dynamic Random Access Memory (DRAM)~\cite{dennard1968field, dennard1974design}
is the dominant main memory technology across a broad range of computing systems
because of its high capacity at low cost~\cite{keeth2007dram, markoff2019ibm,
nature2018memory, ibm2021dram}. \amb{\ame{B}uilding modern DRAM chips requires
specialized design and manufacturing techniques (e.g., custom process
nodes~\cite{kim1999assessing} \amc{and bespoke} materials~\cite{kittl2009high})
\amf{developed across more than half a century of rich
history~\cite{ibm2023icons}}, so \amc{the computing industry employs a
\emph{separation of concerns} to explicitly divide responsibilities between
\emph{DRAM producers} (e.g., manufacturers) and \emph{DRAM consumers} (e.g.,
cloud architects, processor and system-on-a-chip designers, memory module
vendors, etc.).} DRAM producers typically develop highly-optimized DRAM
chips as standalone mass-market components \amc{with clearly-specified
interfaces, independently addressing design concerns related to DRAM
technology.} DRAM consumers then integrate these DRAM chips to develop a broad
range of DRAM-based systems. \amf{This approach enables both parties to specialize
their designs and preserve trade secrets while working around a common interface.}}

\amf{In the last decade, however, \amb{this long-standing separation of concerns
has been challenged by \amf{worsening DRAM technology scaling
difficulties that manifest in two problematic ways.}}}

\amf{\amb{First, improvements to DRAM \ami{access latency} and storage capacity
\ami{are slowing down due to DRAM technology scaling
challenges}~\cite{mutlu2015main, ieee2022more}.} To illustrate this problem, we
survey \ami{58 publicly-available} DRAM chip datasheets\footnote{\amf{We conduct
a best-effort survey \amh{of publicly-available datasheets. We} itemize the
specific datasheets we reference in Appendix~\ref{position:appendix_b}.}} in
Section~\ref{newsubsec:slowdown}, broadly sampling chips \ami{from across 19
DRAM manufacturers spanning} the past five decades. We study the evolution of
access latency and storage capacity characteristics, showing slowdowns in
improvements to both metrics in the past ten to twenty years. For example,
average annual improvements to the data access latency (governed by the timing
parameter \trcd{}) reduced by 69.5\% (from 2.66\% to 0.81\%) before and after the
year 2000, with little to no improvement in the past decade. This is consistent
with similar \amh{surveys conducted by} prior works~\cite{son2013reducing,
chang2017thesis, lee2013tiered, hennessy2011computer, chang2016understanding,
lee2016reducing, isaac2008remarkable, choi2015multiple, borkar2011future,
nguyen2018nonblocking, lee2015adaptive}.}

\amf{Second,} \ame{conventional approaches to managing scaling challenges used
by DRAM producers (e.g., in-DRAM mechanisms to mitigate worsening memory errors)
no longer suffice to hide the underlying problems from DRAM consumers. For
example, memory errors caused by the RowHammer vulnerability \amf{are a serious
and growing threat to system-level \ami{robustness (i.e., security, reliability,
and safety)}~\cite{frigo2020trrespass, pessl2016drama, mutlu2017rowhammer,
mutlu2019rowhammer, mutlu2023fundamentally}. \ami{Experimental
studies~\cite{kim2020revisiting, kim2014flipping, park2016statistical,
park2016experiments, hassan2021uncovering} throughout the past decade
demonstrate that these errors can be consistently induced across a broad range
of DRAM chips. In particular, Kim et al.~\cite{kim2020revisiting} show that
RowHammer errors can be induced much more quickly in modern chips, i.e., with
only 4.8K memory activations for chips from 2019--2020, which is 14.4$\times$
lower than the 69.2K activations required for older DRAM chips from 2010--2013.}
Section~\ref{newsec:dram_tech_scaling} further discuss how} DRAM technology
design concerns that were previously hidden by the separation are now breaking
through to become consumer-facing problems that impact system-wide
\ami{performance and robustness}.}

\ame{We believe that overcoming the DRAM scaling problem requires creative,
holistic thinking from \emph{everyone} involved, including both producers and
consumers throughout industry and academia. Unfortunately, we observe two key
limitations with today's separation of concerns that discourage progress toward
addressing scaling challenges:}

\begin{enumerate}
    \item \amb{The \emph{industry-wide DRAM standards} that \ami{specify how to
    implement} the separation (e.g., JEDEC DDR\emph{n}~\cite{jedec2012ddr4,
    jedec2020ddr5}, \ami{HBM\emph{x}~\cite{jedec2021high,
    jedec2022high}})\footnote{\amb{DRAM standards specify a DRAM chip's
    \ami{microarchitecture}, including its interface, configuration, and performance
    characteristics as visible to DRAM consumers.
    Section~\ref{newsec:bg_standards} explains standards in detail.}} do so
    \emph{imperfectly}, requiring laborious revision to adapt to failures in
    separation.}

    \item \amb{The existing separation is \emph{too strict}, \ame{which
    constrains each party's solution space and stifles} opportunities to explore
    new \ami{ways to address} the scaling challenges.}
\end{enumerate}

\amb{These observations stem from a combination of two key sources of evidence.}

\amb{\amd{First}, recent \ami{robustness} challenges caused by memory errors
\emph{have already broken}} the separation of concerns established by current
DRAM standards. Section~\ref{newsubsec:breakdown} references two specific
instances of this problem, RowHammer~\cite{kim2014flipping, bains2014row,
mutlu2017rowhammer, mutlu2019rowhammer, mutlu2023fundamentally} and on-die error
correction~\cite{son2015cidra, kwak2017a, patel2020bit, gong2018duo}, showing
that both cases \amb{expose memory errors caused by DRAM technology behavior in
a way that is undefined by existing DRAM standards. Although recent
\amf{changes} to the standards~\cite{jedec2021system, jedec2021near,
jedec2020lpddr5, jedec2020ddr5} \amf{discuss these problems and provide limited
solutions}, undefined chip behavior remains a serious and worsening problem for
DRAM consumers~\cite{qureshi2021rethinking, mutlu2023fundamentally,
saroiu2022price, saroiu2022configure, jattke2022blacksmith}.}

\amb{\amd{Second, many promising approaches to address DRAM scaling challenges
in today's chips~\cite{baek2014refresh, brasser2017can,
cardarilli2000development, chandrasekar2014exploiting, chang2016understanding,
chen2013e3cc, chen2015ecc, gao2019computedram, ghosh2007smart,
greenfield2012throttling, hajinazar2021simdram, hassan2016chargecache,
hwang2012cosmic, jafri2020refresh, jian2013adaptive, jian2013low,
katayama1999fault, khan2014efficacy, khan2016case, khan2016parbor,
khan2017detecting, kim2000dynamic, kim2003block, kim2014flipping, kim2015bamboo,
kim2015frugal, kim2018dram, kim2018solar, kim2019d, konoth2018zebram,
koppula2019eden, lee2015adaptive, lee2017design, lin2012secret, liu2012raidr,
manzhosov2021muse, mathew2017using, mcelog2021bad, meza2015revisiting,
mutlu2018rowhammer, nair2013archshield, nair2016xed, nvidia2020dynamic,
ohsawa1998optimizing, olgun2021pidram, olgun2021quac, patel2017reach,
patil2021dve, qureshi2015avatar, saileshwar2022randomized, seshadri2013rowclone,
seshadri2015fast, seshadri2016buddy, seshadri2017ambit, seshadri2019dram,
udipi2012lot, van2018guardion, venkatesan2006retention, wang2014proactivedram,
wang2018content, wang2018reducing, yaglikci2020blockhammer, yoon2010virtualized,
zhang2016restore, zhang2021quantifying} rely upon exploiting the benefits of
\emph{deliberately breaking} the separation of concerns. These approaches employ
\emph{system-memory \amh{cooperation}}~\cite{mutlu2013memory, mutlu2014research,
mutlu2015main, mutlu2023fundamentally, loughlin2022software}, demonstrating
significant system-level benefits from implementing mechanisms outside the DRAM
chip to supplement on-chip solutions built by DRAM producers.
Section~\ref{newsec:overcoming_scaling} surveys these proposals categorized by
the particular DRAM scaling challenge they tackle: (1) improving memory access
latencies (Section~\ref{newsec:case_perf}); (2) reducing DRAM refresh overheads
(Section~\ref{newsec:case_pwr}); (3) \amf{securely} defending against the
RowHammer vulnerability (Section~\ref{newsec:case_sec}); and (4) addressing
worsening memory errors (Section~\ref{newsec:case_rela}). Unfortunately, we
observe that today's separation of concerns does not support producers and
consumers to adopt these methods \ami{with ease}. Instead, doing so requires
them to \emph{work around} DRAM standards, which is}} impractical for the
overwhelming majority of consumers due to the risks and costs inherent in custom
modifications to DRAM chips and unstandardized behavior.
 
\amc{Based on these observations, we conclude that both the separation of
concerns and the standards that \ami{specify} them are outdated for today's DRAM
landscape.} \amb{To rethink the separation of concerns in a modern context, we
refer to \amd{each of the four cases of system-memory \amh{cooperation} that we
study in Section~\ref{newsec:overcoming_scaling}. In each case, we review
\emph{how} prior proposals break the separation of concerns so that we can
better understand its limitations today.}} 

\amb{We find that the single most important barrier to advancement in all four
cases is the \emph{consumer's lack of insight into DRAM reliability.} \amg{For
example, Section~\ref{newsec:case_sec} explains how knowing certain properties
of memory errors (e.g., correlation with physical chip locations, memory access
patterns, and operating parameters) is essential for developing secure defenses
against the RowHammer vulnerability.} The existing separation of concerns
effectively abstracts details of a DRAM chip's internal operation \amc{away from
consumers}} \amd{to the extent that consumers do not have the necessary context
to \ami{properly} reason about \ami{and evaluate} how operating the chip in a
particular way will impact its reliable operation. This encompasses operating
points both within and outside of manufacturer recommendations; in either case,
the consumer lacks the context necessary to accurately determine how their
design decisions outside of the DRAM chip (e.g., \amh{in the} memory controller)
affect the DRAM chip's reliable operation.}

\amd{To gain further insight into this problem,} \amf{we study the general
process of memory reliability testing in
Section~\ref{newsec:dram_modeling_and_testing}. Our analysis suggests that the
entire \amh{testing} process is grounded on knowing relevant \ami{properties} of
a DRAM chip's microarchitectural design, such as the physical organization of
cells within the storage array and how they encode data at a circuit level.
Using this information, a DRAM consumer can build models and test methodologies
to explore the full design space surrounding commodity DRAM chips.}

\amb{Based on our analysis, we advocate \amf{revising} both the separation of
concerns and the standards that \ami{specify} them to incorporate \emph{limited
information transparency} between DRAM producers and consumers.} In particular,
explicitly communicating basic DRAM design and test characteristics from DRAM
producers to consumers empowers the consumer with the context to understand how
different system-level design choices (e.g., optimizations in the processor and
memory controller) will affect DRAM chip operation.
\amc{Section~\ref{position:subsec:what_to_release} \amf{identifies information
to communicate \ami{by drawing on examples from research studies}},} including
(1) basic microarchitectural \ami{properties} (e.g., organization of physical
rows, sizes of internal storage arrays) and (2) best practice \mpf{guidelines
for reliability testing} (e.g., test patterns for key error
mechanisms).\footnote{\amc{Consumers with access to appropriate testing
infrastructure can reverse-engineer much of this
information}~\cite{jung2016reverse, lee2015adaptive, patel2019understanding,
chang2016understanding, mukhanov2020dstress, kim2018dram, kraft2018improving,
kim2018solar, orosa2021deeper, hassan2021uncovering, lee2017design}\amc{.
\ami{However, existing techniques may not reveal all possible necessary
information (e.g., due to inaccessible components such as remapped rows)}.}}
\amc{Section~\ref{sec:spec_as_problem} further explains how} access to this
information provides \amc{a practical degree of insight \ami{into DRAM
operation} for DRAM consumers to work with, without \amf{compromising} DRAM
producers' trade secrets \ami{or} the cost advantages of commodity DRAM.}

\amb{\amc{We} advocate for \amc{industry} to incorporate} information
transparency into DRAM standards \amc{through} a two-step approach involving all
DRAM stakeholders, including producers and consumers.

\amc{\emph{Step 1: Early Adoption.}} \amc{Initially, we recommend conceptually
revising the separation of concerns without yet revising DRAM standards. This
step targets} DRAM chips already in the field, \amc{asking both DRAM producers
and consumers to \amf{voluntarily} release information they already have at
hand. We propose each party to take a different approach as follows:}
\begin{itemize}
    \item \emph{Consumers:} \amc{Contribute to a crowdsourced database of information
    obtained through third-party testing of commodity DRAM chips on the market
    (e.g., as conducted in numerous studies discussed in Section~\ref{newsec:overcoming_scaling}).}
    \item \emph{Producers:} \amc{Publish} information (e.g., using datasheet
    revisions, \amc{whitepapers,} or online resources) about their products,
    possibly limited to information that they already have on hand from past
    records (i.e., \amc{information} that requires minimal logistical effort to
    release).
\end{itemize}
\amc{Through these two avenues, \emph{all} DRAM \mpf{stakeholders will} benefit
from the release of information:} \amf{consumers will gain access to a wider
solution space for addressing DRAM scaling challenges, and producers will gain
access to insights and new solution directions from consumers' efforts. In
particular, these benefits require \emph{neither} changes} to existing DRAM
hardware or standards (though standardizing the information release could
streamline the process) \amf{nor forced disclosure of sensitive information by
either party}.

\amc{\emph{Step 2: Long-Term Revision.} In the long run, we} propose
\amc{revising} DRAM standards \amc{to include industry-standard information,
tools, and specifications for a wider range of DRAM operating points. In
addition to the information requested by Step 1, we identify two recommendations
that our studies show would be beneficial for consumers:} (1) reliability
guarantees for how a chip is expected to behave under certain operating
conditions (e.g., predictable behavior of faults~\cite{criss2020improving}); (2)
disclosure of industry-validated DRAM reliability models and testing strategies
suitable for commodity DRAM chips (e.g., similar to how JEDEC
JEP122~\cite{jedec2016failure}, JESD218~\cite{jedec2010ssdrequirements}, and
JESD219~\cite{jedec2010ssdendurance} address Flash-memory-specific error
mechanisms~\cite{cai2017error, cai2012error, cai2018errors} such as
floating-gate data retention~\cite{cai2015data, luo2018heatwatch,
luo2018improving, cai2012flash} and models for physical phenomena such as
threshold voltage distributions~\cite{cai2013threshold, cai2013program,
cai2015read, luo2016enabling, cai2017vulnerabilities, cai2014neighbor}).
\amc{Revising DRAM standards in this way will align the standards with a more
permissive separation of concerns going forward, thereby encouraging cooperation
between producers and consumers in pursuit of building \ami{more robust and
higher-performance} future computing systems.}

We make the following contributions:

\begin{itemize} 
    \item \amc{We make a case to rethink the \amh{long-standing} separation of
    concerns between DRAM producers and consumers \amh{to enable and encourage
    new solutions for addressing worsening} DRAM technology scaling challenges.}

    \item \amc{We \amg{motivate and support} our case \amh{by thoroughly
    reviewing} prior work, including \amh{(1) four case studies that survey
    system-memory cooperative techniques to address DRAM scaling challenges and
    (2) a historical survey of \ami{DRAM chip capacity, latency, and energy}
    characteristics based on datasheets.} We
    open-source~\cite{datasheetsurveygithub} our dataset.}

    
    \item \amh{We provide a new perspective on memory reliability testing from
    the viewpoint of DRAM consumers, identifying access to a DRAM chip's
    \ami{microarchitectural details} \amg{as both a key \ami{challenge} and
    enabler for \amh{system-memory} cooperative solutions to DRAM scaling
    challenges.}}

    \item \amc{We propose a \amh{practical plan to encourage new solutions to
    modern DRAM scaling challenges by revising} both the separation of concerns
    and \ami{how it is specified by DRAM standards today}.}
\end{itemize} 
\section{DRAM Standards as a Separation of Concerns}
\label{newsec:bg_standards}

\amf{Industry-wide DRAM standards split the responsibilities of \emph{building}
and \emph{integrating} commodity DRAM chips to DRAM producers and consumers,
respectively. This section reviews how standards achieve this and its
implications for DRAM producers and consumers.}

\subsection{DRAM Standards}

\amf{DRAM standards} carefully balance the needs of both \amf{producers and
consumers through industry-wide consensus. Today,} the JEDEC
consortium~\cite{jedec2021jc42} maintains a limited set of standards describing
commodity DRAM chips with different target applications, e.g., general-purpose
DDR\emph{n}~\cite{jedec2008ddr3, jedec2012ddr4, jedec2020ddr5},
\mpk{bandwidth-optimized HBM\emph{n}~\cite{jedec2021high, jedec2022high},}
mobile-oriented LPDDR\emph{n}~\cite{jedec2014lpddr4, jedec2020lpddr5},
graphics-oriented GDDR\emph{n}~\cite{jedec2016gddr5, jedec2016gddr6}.

\amf{DRAM standards specify all aspects of a DRAM chip's design that pertain to
the interface between producers and consumers, including the chip's access
interfaces, configuration mechanisms,} and performance characteristics. \amf{By
doing so, standards} effectively abstract DRAM technology-level concerns (e.g.,
cell-to-cell variation, reliability challenges) into predictable usage patterns,
thereby reducing a complex storage technology into a modular computing
component. 

\subsection{Advantages of Standardized DRAM}

\amf{Standardized DRAM enables the widespread use of highly-optimized DRAM
chips. This is because standards scope DRAM technology to a few \amh{fixed
components (i.e., standardized chips) with clearly-defined interfaces} that both
producers and consumers can optimize towards without concerning themselves with
the other party's design challenges. This gives each party the freedom to
explore different solutions to DRAM design challenges based on their design
goals. For example, empirical studies of DRAM chips~\cite{beigi2023systematic,
frigo2020trrespass, jattke2022blacksmith, orosa2021deeper, patel2020bit} show
that different DRAM producers mitigate memory errors using a broad range of
different on-chip error mitigation techniques. In general, producers are free to
innovate in any way that does not violate the specifications established by
standards, and consumers can build upon those specifications in any way to meet
their systems' needs.}

\subsubsection{DRAM Producers' Trade Secrets}

\amh{DRAM producers closely guard their innovations because trade secrets are a
key component of business competitiveness and}
\amf{success~\cite{nair2013archshield, gong2017dram, childers2015achieving,
cost1997yield}. Producers who build standards-compliant chips are competing in a
commodity market, so they} seek profitability though economies of
scale~\cite{kang2010study, dell1997white, lee2013strategic, croswell2000model}.
\amf{Each producer develops and uses home-grown, highly-optimized design,
manufacturing, and testing processes that amortize costs in high volume
production, thereby \amh{maximizing} per-chip profit margins.}

\amf{As a result, DRAM producers publish only \amh{what} information DRAM
standards require, such as access timing specifications and circuit
characteristics needed for chip integration. Additional \amh{information} not
specified by standards (e.g., internal circuit designs, chip error rates)
\amh{is} kept in-house. Although \amh{such} details can \amh{often} be inferred
through reverse-engineer\amh{ing} studies~\cite{patel2017reach,
kraft2018improving, liu2013experimental, patel2020bit, kim2020revisiting,
jung2016reverse, barenghi2018software, hassan2021uncovering, farmani2021rhat,
frigo2020trrespass, wang2020dramdig, pessl2016drama, jiang2021trrscope,
kim2018solar, chang2016understanding, lee2017design, nam2023x} and chip
teardowns~\cite{james2010silicon, torrance2009state} (discussed further in
Section~\ref{newsec:overcoming_scaling}), producers have no obligation to
communicate this information to consumers.}

\subsection{Using Non-Commodity DRAM Chips}
\label{subsec:non_commodity_chips}

\amf{Consumers can use non-commodity DRAM chips by either (i) privately working
with DRAM producers to build customized chips or (ii) buying specialized or
otherwise domain-optimized DRAM chips (e.g., high
reliability~\cite{smart2021rugged, im2016im}, low
latency~\cite{micron2021rldram}). These chips are generally still compliant with
JEDEC standards, though they may provide additional unstandardized features.}

\amf{Developing and using non-commodity DRAM can benefit industry-wide
standards. For example, the HBM standard (JESD235~\cite{jedec2021high}) largely
grew from private industry collaborations between AMD and SK
Hynix~\cite{macri2015amd}. Similarly, features pioneered in non-commodity DRAM
can be integrated into newer standards.}

\amf{Unfortunately, \amh{innovating through} non-commodity DRAM is \amh{a} slow
and costly \amh{process because it} forgoes the advantages of mainstream DRAM
chips. \amh{Non-commodity chips are} typically feasible only for consumers in
specific industries (e.g., imaging, networking~\cite{intel2023rldram}) or with}
significant stake in the global DRAM market (e.g., large-scale cloud vendors).
\amf{Our work takes inspiration from non-commodity DRAM to enable \emph{all}
consumers to pursue such innovations, ultimately benefitting the DRAM industry
as a whole.}

\subsection{Creating or Modifying DRAM Standards}
\label{newsubsec:modifying_dram_standards}

\amf{Changes to DRAM standards require participation from all stakeholders
throughout the DRAM industry, including JEDEC personnel, DRAM producers, and
DRAM consumers. Therefore, making changes is a slow process that can involve
non-technical elements, such as political motivations and business
goals~\cite{church2007strategic}. Major changes typically follow one of three
different paths. First, a standards committee comprising experts from all
stakeholders may directly draft a new standard. Second, new standards may grow
out of the development and use of non-commodity DRAM as discussed in
Section~\ref{subsec:non_commodity_chips}. Third, existing standards may be
updated or supplemented by JEDEC committees for special issues, such as
LPDDR4X~\cite{jedec2021lpddr4x}, 3D-stacked DRAM~\cite{jedec20213ds}, and
RowHammer~\cite{jedec2021system, jedec2021near}.}

\section{Challenges of DRAM Technology Scaling}
\label{newsec:dram_tech_scaling}

\amf{DRAM's primary competitive advantage is its low
cost-per-capacity~\cite{keeth2007dram, markoff2019ibm, nature2018memory,
ibm2021dram}, which DRAM \amb{producers} maintain by continually improving chip}
storage densities. \amf{Doing so across successive product generations requires
carefully balancing shrinking} physical feature sizes, optimizing circuit areas,
and mitigating \amf{worsening} memory errors~\cite{kang2014co, cha2017defect,
nair2013archshield, micron2017whitepaper, park2015technology, son2015cidra,
micron2022quaterly}.

\amf{Unfortunately, today's DRAM faces two key challenges to continued
technology scaling: (i) the slowdown of generational improvements to storage
capacity, \ami{access latency}, and power consumption~\cite{chang2017thesis,
lee2016reducing, ghose2018your}; and (ii) the breakdown of conventional
approaches to mitigate memory errors. This section reviews both challenges in
detail.}

\subsection{Slowdown of Generational Improvements}
\label{newsubsec:slowdown}

\amf{Advancements in DRAM chip storage density have been central to increasing
demands for memory capacity since the inception of DRAM
technology~\cite{ibm2023icons, akarvardar2023technology}. Today's emerging
data-intensive applications and systems in domains such as AI, cloud, and HPC
continue to demand greater memory capacity at an unprecedented
scale~\cite{van2012role, ousterhout2010case, lee2021reliability, sun2022rm,
park2022scaling}. Unfortunately, technology shrinkage throughout the past two
decades has yielded diminishing benefits for chip storage capacity, access
latency, and refresh overheads because of the growing costs and overheads of
maintaining reliable chip operation at smaller technology node
sizes~\cite{ieee2022more, mutlu2015main}.}

\amg{To better understand this slowdown of generational improvements, we survey
manufacturer-reported DRAM chip capacities, access timings, and current
consumption characteristics given by} 58 publicly-available DRAM chip datasheets
from across 19 different DRAM manufacturers with datasheet publication dates
between 1970 and 2021.\footnote{This data encompasses DRAM chips from both asynchronous
(e.g., page mode, extended data out) and synchronous (e.g., SDRAM, DDR\emph{n})
DRAM chips. Appendix~\ref{position:appendix_a} describes our data collection
methodology in further detail, and Appendix~\ref{position:appendix_b} provides
an overview of our dataset, which is publicly available on
GitHub~\cite{datasheetsurveygithub}.} \amg{The remainder of this section
individually analyzes chip capacity, access timings, and refresh overheads in
the context of our survey.}

\subsubsection{\amd{Chip Storage Capacity}}
\label{subsubsec:mot_storage_capacity}

\begin{figure}[ht]
\begin{minipage}{\linewidth}
\renewcommand{\thempfootnote}{\roman{mpfootnote}}
    \centering
    \includegraphics[width=\linewidth]{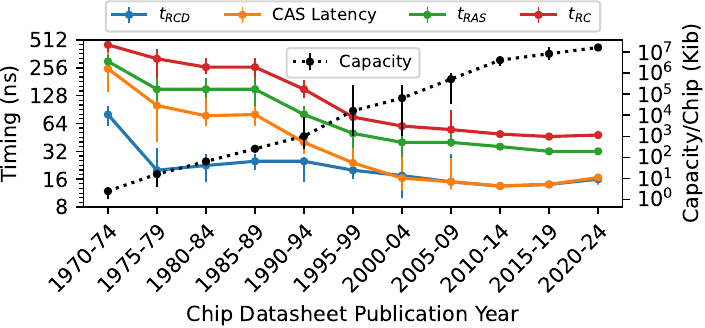}
    \caption[]{%
    Semi-log plot showing the evolution of key DRAM access timings (left) and per-chip storage capacity (right)$^\mathrm{i}$ across each 5-year period of time.}
    \label{fig:timings_idds}
    {\footnotesize $^\mathrm{i}$JEDEC-standardized parameters\cite{jedec2020ddr5} found in DRAM chip datasheets:}\\
    \indent\indent\indent
    \begingroup%
    \scriptsize
    \begin{tabular}{ll}%
        \textbf{Parameter} & \textbf{Definition}\\\hline
        \trcd{} & minimum row activation to column operation delay\\
        CAS Latency & read operation to data access latency\\
        \tras{} & minimum row activation to precharge delay\\
        \trc{} & minimum delay between accesses to different rows
    \end{tabular}%
    \endgroup%
\end{minipage}
\end{figure}

Figure~\ref{fig:timings_idds} shows \amf{the time evolution of} per-chip storage
capacity and four key DRAM operating timings \mpn{(all shown in log scale)}. We
observe \amg{that storage capacity has grown exponentially over time alongside
improvements to} all four timing parameters \amf{(timings are discussed in
Section~\ref{subsubsec:mot_access_latency}). However, storage capacity
\amg{growth has} slowed down markedly since 2010, dropping from an exponential
growth factor of 0.341 per year for 1970-2000 to 0.278 for 2000-2020. \amg{This
is consistent with recent challenges in scaling beyond 16~Gb chip densities, and
this slowdown is expected to continue going forward~\cite{gervasi2019will,
ieee2022more}.}}

\noindent

\subsubsection{\amd{Chip Access Latency}}
\label{subsubsec:mot_access_latency}

DRAM access latency has not significantly improved relative to storage capacity
over the last two decades~\cite{son2013reducing, chang2017thesis, lee2013tiered,
chang2016understanding, lee2016reducing, isaac2008remarkable, choi2015multiple,
borkar2011future, nguyen2018nonblocking}. This makes DRAM an increasingly
significant system performance bottleneck today, especially for workloads with
large footprints that are sensitive to DRAM access
latency~\cite{hsieh2016accelerating, ferdman2012clearing, gutierrez2011full,
hestness2014comparative, huang2014moby, zhu2015microarchitectural,
oliveira2021damov, boroumand2018google, boroumand2021google, koppula2019eden,
kanellopoulos2019smash, son2013reducing, mutlu2013memory, wilkes2001memory,
wulf1995hitting, mutlu2007stall, mutlu2003runahead, kanev2015profiling,
mutlu2014research, bera2019dspatch, bera2021pythia, liu2019binary,
ghose2019processing, shin2014nuat, ghose2019demystifying, gomez2021benchmarking,
gomez2021benchmarkingmemory, giannoula2022towards}. Therefore, there is
significant opportunity for improving overall system performance by reducing the
memory access latency~\mpk{\cite{chandrasekar2014exploiting,
chang2016understanding, kim2018solar, lee2015adaptive, lee2017design,
wang2018reducing, zhang2016restore, hassan2016chargecache, koppula2019eden,
mathew2017using, zhang2021quantifying, kim2020improving, lee2016simultaneous}}.
Although \mpk{conventional} latency-hiding techniques (e.g., caching,
prefetching, multithreading) can potentially help mitigate many of the
performance concerns, these techniques (1) fundamentally do not change the
latency of each memory access and (2) fail to work in many cases (e.g.,
irregular memory access patterns, random accesses, huge memory footprints).

\amf{Figure~\ref{fig:timings_idds} shows that \emph{none} of the
four timings we study} have improved significantly in the last two decades.
For example, the median \trcd{}/CAS Latency/\tras{}/\trc{} reduced by
2.66/3.11/2.89/2.89\% per year on average between 1970 and 2000, but only
0.81/0.97/1.33/1.53\% between 2000 and 2015.\footnote{We report 2015 instead of
2020 because 2020 shows a regression in CAS latency due to first-generation DDR5
chips, which we believe is not representative because of its immature
technology.} This is consistent with similar \mpj{studies} in prior
work~\cite{son2013reducing, chang2017thesis, lee2013tiered,
hennessy2011computer, chang2016understanding, lee2016reducing,
isaac2008remarkable, choi2015multiple, borkar2011future, nguyen2018nonblocking}.

\subsubsection{Worsening Refresh Overheads}
\label{subsubsec:mot_dram_refresh}

\amf{The circuits that DRAM uses to store data are inherently susceptible to a
wide range of different leakage mechanisms (e.g., capacitor charge leakage),
which ultimately cause data loss if ignored. To prevent this, DRAM standards
mandate} periodic \emph{refresh} operations that intermittently \amf{restore
data values throughout the entire DRAM chip.} Unfortunately, DRAM refresh
\amf{incurs} significant system performance and power
overheads~\mpm{\cite{ohsawa1998optimizing, kim2000dynamic, laudon2006ultrasparc,
liu2012raidr, nair2013archshield, bhati2015flexible, wang2014proactivedram,
baek2014refresh, mathew2017using, bhati2016dram}}, sacrificing almost half of
the total memory throughput and \amf{consuming} \mpk{almost} half of the total
DRAM power for projected 64 Gb chips~\cite{liu2012raidr}.

\begin{figure}[h]
    \centering
    \includegraphics[width=\linewidth]{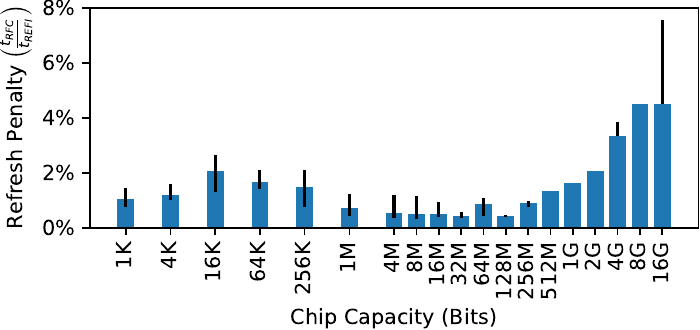}
    \caption[]{\mpm{Refresh penalty} (computed as the ratio between \trfc{} and \trefi{}) for DRAM chips of different storage capacities.}
    \label{fig:da_ref_overhead}
\end{figure}

Figure~\ref{fig:da_ref_overhead} illustrates the performance overheads of DRAM
refresh across the different DRAM chips in our datasheet survey. The data shows
the \emph{refresh penalty},\footnote{\amf{Also referred to as refresh
overhead~\cite{zhang2014cream} and refresh duty cycle~\cite{nair2014refresh}.}}
which is defined as the ratio of two key timing parameters used to govern
refresh operations: \emph{\trfc{}}, the duration of each refresh command, and
\emph{\trefi{}}, the time between consecutive refresh commands. The refresh penalty
represents the average time that a DRAM rank (or bank) is unavailable for access
due to refresh operations~\cite{utah2013dram, stuecheli2010elastic,
balasubramonian2019innovations, cheng2019retention, zhang2014cream}.

We observe that the refresh penalty worsens from a median of 1.04\% for 1~Kib
chips to 2.05\% for 16~Kib chips, then improves to 0.43\% for 128~Mib chips, and
finally worsens to a median of 4.48\% (worst-case of 7.56\% for DDR5 chips) for
16~Gib chips.\mpl{This non-monotonic trend is due to the relative \amh{rates of}
improvement in DRAM access latency and storage capacity: \amf{access (and
therefore, refresh) latencies stagnated around the introduction of $128$~Mib
chips while capacity improvements did not. This data is consistent with both (i)
our more detailed analysis of DRAM refresh timings in
Appendix~\ref{appendix_a:subsec:refresh_timings} and (ii)} trends observed in
prior work~\mpm{\cite{nguyen2018nonblocking, liu2012raidr, bhati2015flexible,
nair2013case, chang2014improving, qureshi2015avatar}}, which expect that future,
higher-capacity DRAM chips will spend an even larger proportion of time
refreshing unless the DRAM refresh algorithm \amh{and its implementation are}
improved.}

\subsection{Breakdown of the Separation of Concerns}
\label{newsubsec:breakdown}

\amd{DRAM is susceptible to a variety of error mechanisms that \amf{worsen with
technology scaling and} can impact system-wide reliability if not carefully
managed. Today's separation of concerns largely \amf{puts the burden to address
these errors on DRAM producers, i.e., to hide them from consumers and provide
the illusion of an error-free memory chip.} Unfortunately, two classes of
\amf{scaling-related errors are} breaking through the separation to impact
consumers \amf{regardless}: random, single-bit errors and RowHammer errors. This
section discusses \amf{these errors and how they break} the separation of
concerns.}

\subsubsection{Worsening Reliability}
\label{subsubsec:mot_rela}

\amg{DRAM suffers from a range of circuit-level error mechanisms that can impact
consumers (e.g., data loss, system failure) if
mismanaged~\cite{horiguchi2011nanoscale, sridharan2012study, schroeder2009dram,
may1979alpha, o1996field, meza2015revisiting, sridharan2015memory}. To mitigate
these errors on-chip, DRAM producers typically use} a combination of
conservative \amg{operating timings} (e.g., \amg{with added safety margins}) and
simple error-correcting hardware within the DRAM chip (e.g., \amg{manufacturing
spare rows and columns to replace faulty ones~\cite{horiguchi2011nanoscale,
jedec2012ddr4, jedec2020ddr5, kim2016ecc,
wada2004post}}).\footnote{\amg{Enterprise-class computing providers (e.g.,
cloud, HPC) may use supplementary error-mitigation mechanisms discussed further
in Section~\ref{newsubsubsec:coop_solutions_for_reliability}}.} These low-cost
techniques enable DRAM producers to provide the illusion of a fully reliable
DRAM chip, \amg{thereby preserving the separation of concerns between producers
and consumers.}

\amg{In contrast, modern DRAM chips exhibit much higher error rates because
technology scaling exacerbates the underlying circuit-level error mechanisms
that cause errors~\cite{gong2017dram, nair2013archshield, beigi2023systematic,
gurumurthi2021hbm3, cha2017defect, kang2014co, lee2022ecmo}. To combat these
errors, DRAM producers use stronger error-mitigation mechanisms in modern DRAM
chips (e.g., on-die ECC~\cite{son2015cidra, gong2018duo, nair2016xed,
jeong2020pair, cha2017defect, pae2021minimal, criss2020improving,
luo2014characterizing, gurumurthi2021hbm3, patel2019understanding, patel2020bit,
patel2021harp, patel2021enabling, jedec2020ddr5}, post-package
repair~\cite{horiguchi2011nanoscale, jedec2012ddr4, jedec2020ddr5, kim2016ecc,
wada2004post}, target row refresh~\cite{saroiu2022price, qureshi2021rethinking,
hassan2021uncovering, frigo2020trrespass}, refresh
management~\cite{jedec2020ddr5, saroiu2022configure}), \amh{which are more
expensive and incur higher performance and energy overheads}.}

\amg{Unfortunately, worsening memory reliability remains a serious problem for
DRAM consumers, especially high-volume consumers for whom \amh{even} modest chip
error rates \amh{are} significant at scale~\cite{sridharan2015memory,
beigi2023systematic}. Although stronger in-DRAM error mitigations are effective
against growing error rates~\cite{beigi2023systematic, gurumurthi2021hbm3}, they
introduce new overheads and challenges for consumers. For example, neither
on-die ECC nor target row refresh correct \emph{all} errors, and the remaining
errors (e.g., uncorrectable errors) \amh{are difficult for consumers to predict
and mitigate because their manifestation depends on the particular} on-die ECC
and/or TRR mechanism used by a given chip~\cite{son2015cidra,
gong2018duo,patel2019understanding, patel2020bit, patel2021harp,
saroiu2022price, frigo2020trrespass, hassan2021uncovering, loughlin2021stop,
farmani2021rhat}. As a result, DRAM consumers face errors that are growing in
both magnitude and complexity, making reliability a key design concern for
continued DRAM scaling.}

\subsubsection{The RowHammer Vulnerability}
\label{subsubsec:mot_dram_security}

RowHammer~\mpk{\cite{kim2014flipping, bains2014row, mutlu2017rowhammer,
mutlu2019rowhammer, yang2019trap, greenfield2014method, mutlu2023retrospective}}
is a scaling-related read-disturb phenomenon affecting modern DRAM chips in
which memory accesses to a given \amg{physical} memory location can induce
bit-flips at other locations. \amg{Significant work~\cite{seaborn2015exploiting,
van2016drammer, gruss2016rowhammer, razavi2016flip, pessl2016drama, xiao2016one,
bosman2016dedup, bhattacharya2016curious, qiao2016new, jang2017sgx, aga2017good,
mutlu2017rowhammer, tatar2018defeating, gruss2018another, lipp2020nethammer,
van2018guardion, frigo2018grand, cojocar2019eccploit,  ji2019pinpoint,
mutlu2019rowhammer, hong2019terminal, kwong2020rambleed, frigo2020trrespass,
cojocar2020rowhammer, weissman2020jackhammer, zhang2020pthammer,
yao2020deephammer, kogler2022half, de2021smash} shows that RowHammer is a
security threat that can be exploited to exfiltrate sensitive data, escalate
user privileges, or otherwise compromise a system. Because RowHammer is
fundamental to DRAM circuit designs, it is a problem for all DRAM-based systems
and becomes significantly worse with continued process technology
scaling~\cite{kim2014flipping, kim2020revisiting, mutlu2019rowhammer,
mutlu2017rowhammer}.} 

\amg{Modern DRAM partially addresses RowHammer through best practices laid out
in DRAM specifications~\cite{jedec2021near, jedec2021system, jedec2020ddr5} and
RowHammer mitigation mechanisms built into DRAM chips by producers, including
\amg{target row refresh (TRR)~\cite{de2021smash, lee2014green,
frigo2020trrespass, hassan2021uncovering, cojocar2019exploiting,
kim2020revisiting, micron20208gb} and refresh management
(RFM)~\cite{saroiu2022configure, jedec2020ddr5}.} Unfortunately, neither
approach suffices to provide fully secure solutions against
RowHammer~\cite{frigo2020trrespass, cojocar2020are, hassan2021uncovering,
jattke2022blacksmith, de2021smash, orosa2021deeper}. Therefore, RowHammer
remains a serious consumer-visible \ami{problem} that challenges the
illusion of a \ami{robust} DRAM chip established by the separation
of concerns.} 
\section{Challenges in Overcoming Scaling Problems}
\label{newsec:overcoming_scaling}

\amh{Developing new solutions to address the DRAM scaling challenges discussed
in Section~\ref{newsec:dram_tech_scaling} requires creative thinking that
today's separation of concerns constrains.} \amg{\amh{This section explores new
techniques based on system-\amh{memory} \amh{cooperation} that holistically
address DRAM scaling challenges with the help of hardware and software
mechanisms at the system level. We review techniques that target each of the
four key DRAM scaling challenges} discussed in
Section~\ref{newsec:dram_tech_scaling}: access latency, refresh overheads,
RowHammer, and worsening reliability. In each case study, we \amh{survey}
relevant prior work to understand how and why today's separation of concerns
discourages system-\amh{memory} cooperative solutions.}

\subsection{Study 1: Long DRAM Access Latency}
\label{newsec:case_perf}

\amd{As Section~\ref{subsubsec:mot_access_latency}
discusses,} slow generational improvements \amf{to} DRAM access latency
\amg{make DRAM a worsening bottleneck for}
latency-sensitive workloads today~\mpm{\cite{hsieh2016accelerating,
ferdman2012clearing, gutierrez2011full, hestness2014comparative, huang2014moby,
zhu2015microarchitectural, oliveira2021damov, boroumand2018google,
boroumand2021google, koppula2019eden, kanellopoulos2019smash, son2013reducing,
mutlu2013memory, wilkes2001memory, wulf1995hitting, mutlu2007stall,
mutlu2003runahead, kanev2015profiling, mutlu2014research, bera2019dspatch,
bera2021pythia, liu2019binary, ghose2019processing, shin2014nuat,
ghose2019demystifying, gomez2021benchmarking, gomez2021benchmarkingmemory,
giannoula2022towards}}. \amg{System-\amh{memory} cooperation promises to
overcome this problem by directly reducing the DRAM access latency beyond what
commodity DRAM chips provide.}

\subsubsection{Solutions to Reduce the DRAM Access Latency}

Prior works have taken two major directions \amd{to improve the DRAM access
latency}. First, many works~\mpk{\cite{chandrasekar2014exploiting,
chang2016understanding, kim2018solar, lee2015adaptive, lee2017design,
wang2018reducing, zhang2016restore, hassan2016chargecache, koppula2019eden,
mathew2017using, zhang2021quantifying}} show that the average DRAM access
latency can be shortened by reducing DRAM access timings for particular memory
locations that can tolerate faster accesses. This can be done safely because,
although DRAM standards call for constant access timings across all memory
locations, the minimum viable access timings that the hardware can support
actually differ \amh{across} memory locations due to factors such as
heterogeneity in the circuit design~\cite{lee2017design, lee2016reducing} and
manufacturing process variation \amh{in} circuit
components~\mpm{\cite{chandrasekar2014exploiting, chang2016understanding,
kim2018solar, chang2017understanding, lee2015adaptive}}.

\xmt{Exploiting these variations in access timings to reduce the average memory
access latency \amh{provides} significant system performance improvement. For
example, Chang et al.~\cite{chang2016understanding} experimentally show that
exploiting access latency variations \amh{provides} an average 8-core system
performance improvement of 13.3\%/17.6\%/19.5\% for real DRAM chips from three
major DRAM manufacturers. Similarly, Kim et al.~\cite{kim2018solar} show that
exploiting access latency variations induced by DRAM sense amplifiers provides
an average (maximum) \amh{4-core} system performance improvement of 4.97\%
(8.79\%) versus using default DRAM access timings based on data obtained from
282 commodity LPDDR4 DRAM chips.}

Second, other works~\mpk{\cite{gao2019computedram, kim2018dram, olgun2021pidram,
olgun2021quac, kim2019d, seshadri2013rowclone, seshadri2015fast,
seshadri2017ambit, seshadri2019dram, hajinazar2021simdram, seshadri2016buddy,
seshadri2017simple, seshadri2020indram, talukder2018exploiting,
talukder2018ldpuf, yue2020dram, hashemian2015robust, schaller2018decay}} show
that commodity DRAM can perform massively-parallel computations (e.g., at the
granularity of an 8 KiB DRAM row) by exploiting \mpk{the underlying analog
behavior of DRAM operations} (e.g., charge sharing between cells). These works
show that such computations can significantly improve overall system performance
and energy-efficiency by both (1) reducing the amount of data transferred
between the processor and DRAM and (2) exploiting the relatively high throughput
of row-granularity operations. For example, Gao et al.~\cite{gao2019computedram}
show that in-DRAM 8-bit vector addition is $9.3\times$ more energy-efficient
than the same computation in the processor, primarily due to avoiding the need
for off-chip data transfers. Similarly, Olgun et al.~\cite{olgun2021pidram}
\amg{experimentally} demonstrate that in-DRAM copy and initialization techniques
can improve the performance of system-level copy and initialization
\amg{operations} by $12.6\times$ and $14.6\times$, respectively.

\subsubsection{\amg{Application to Today's Commodity DRAM Chips}}
\label{subsec:adapting_commodity_dram_lat}
\label{subsec:limitng_assumptions_dram_lat}

\amg{Unfortunately, both reducing DRAM access timings and exploiting DRAM's
massively-parallel analog behavior are discouraged by today's separation of
concerns.} In both cases, \emph{new} DRAM access timings must be \mpk{determined
that} ensure \amg{new or modified DRAM operations} can be performed predictably
and reliably \mpk{ under all \amg{operating} conditions}. 

To identify \mpk{\amg{new} access} timings, \amg{the majority of prior
works}~\mpk{\cite{kim2018dram, kim2019d, talukder2019prelatpuf,
gao2019computedram, olgun2021quac, chang2016understanding,
chang2017understanding, lee2015adaptive, chandrasekar2014exploiting,
ghose2018your, hassan2017softmc, chang2017thesis, lee2016reducing,
david2011memory} perform extensive experimental characterization studies across
many \amg{(e.g., hundreds or thousands of)} DRAM chips.} These studies account
for three primary sources of variation that affect the access timings of a given
memory location. First, process variation introduces random variations between
DRAM chip components (e.g., cells, rows, columns). Second, a manufacturer's
particular circuit design introduces structural variation (called design-induced
variation~\cite{lee2017design}) that deterministically affects access timings
based on a component's location in the overall DRAM design (e.g., cells along
the same bitline~\cite{kim2018solar}, cells at the borders of internal storage
arrays~\cite{lee2017design}). Third, the charge level of a DRAM cell varies over
time due to leakage and the effects of DRAM accesses~\cite{shin2014nuat,
hassan2016chargecache}. Experimentally determining the \mpk{new predictable and
reliable} access timings requires \mpk{properly} accounting for all three
sources of variation \mpk{under all operating conditions}.

\amg{For a typical DRAM consumer, \amh{determining new access timings using a}
\amg{custom DRAM} testing methodology is impractical \amh{without assistance
from DRAM producers}.} Choosing runtime (e.g., data and access patterns) and
environmental (e.g., temperature, voltage) testing conditions \emph{in a
meaningful way} requires understanding the error mechanisms involved in
timing-related errors~\cite{khan2017detecting}, including (but not limited to)
\ami{microarchitectural design details}, such as internal substructure
dimensions (e.g., subarray sizing)~\cite{kim2018solar, lee2017design}, the
correspondence between logical DRAM bus addresses and physical cell
locations~\mpm{\cite{chang2016understanding, lee2015adaptive, khan2016parbor}},
and the order of rows refreshed by each auto-refresh
operation~\cite{shin2014nuat}. \mpk{\amg{Therefore, consumers who lack
trustworthy access to this information} are discouraged from exploring
improvements to the commodity DRAM access latency.}
\subsection{Study 2: DRAM Refresh Overheads}
\label{newsec:case_pwr}

DRAM refresh \amg{overheads are a key} design concern \mpf{in modern systems as
\amg{discussed in Section~\ref{subsubsec:mot_dram_refresh}. \amh{System-memory}
cooperation can overcome this problem by eliminating or otherwise mitigating
unnecessary refresh operations, thereby improving overall system performance and
energy efficiency.}}

\subsubsection{Solutions to Reduce DRAM Refresh Overheads}

Prior works~\mpo{\cite{liu2012raidr, ohsawa1998optimizing,
wang2014proactivedram, venkatesan2006retention, lin2012secret,
nair2013archshield, ghosh2007smart, qureshi2015avatar, khan2014efficacy,
khan2016case, khan2016parbor, khan2017detecting, jafri2020refresh,
kim2000dynamic, kim2003block, katayama1999fault, patel2017reach,
mathew2017using}} \amg{address refresh overheads by exploiting the} fact that
\emph{most} refresh operations are unnecessary.\footnote{Latency-hiding
techniques (e.g, prefetching, memory command scheduling, on-chip caching, etc.)
and parallelization of refresh and access
operations~\mpk{\cite{chang2014improving, nguyen2018nonblocking, pan2019hiding,
stuecheli2010elastic, mukundan2013understanding}} help mitigate performance
overheads but do not change the total number of refresh operations issued.
\mpk{As a result, such techniques \amg{do not reduce refresh energy
consumption}}. These techniques are also imperfect in many cases \mpk{where
latency-hiding is impractical (e.g., row conflicts between refresh and access
commands, larger memory footprints than available caching
resources)~\mpm{\cite{nair2013case, pan2019hiding, zhang2014cream,
chang2014improving}}.}} The \amg{standardized} DRAM refresh algorithm refreshes
all cells frequently (i.e., at the worst-case rate) to simplify DRAM refresh and
guarantee correctness. However, each cell's data retention characteristics vary
significantly due to a combination of data-dependence~\mpk{\cite{khan2016parbor,
khan2017detecting, liu2013experimental, khan2014efficacy, patel2017reach}} and
process variation~\mpo{\cite{hamamoto1995well, hamamoto1998retention,
gong2017dram, nair2013archshield, liu2013experimental, liu2012raidr,
wang2014proactivedram}}. \mpi{As a result, eliminating unnecessary refresh
operations} \amh{provides} significant \mpk{power reduction and performance
improvement.} For example, Liu et al.~\cite{liu2012raidr} demonstrate an average
\mpl{energy-per-access and system performance improvement of 8.3\% and 4.1\%,
respectively, for 4~Gib chips (49.7\% and 107.9\% for 64~Gib chips)} when
relaxing the refresh rate at the row granularity. Therefore, reducing refresh
overheads potentially \amh{benefits all DRAM-based systems}.

\subsubsection{Application to Today's Commodity DRAM Chips}
\label{subsec:adapting_commodity_dram_ref}
\label{subsec:limitng_assumptions_dram_ref}

Reducing unnecessary refresh operations in commodity DRAM chips \amg{comprises}
two key steps. First, the memory controller must reduce the frequency of
periodic refresh operations. This is \mpk{achievable (though not necessarily
supported to arbitrary values) using commodity DRAM chips because the memory
controller manages DRAM refresh timings. For example, the memory controller
might} relax the rate at which it issues refresh operations \mpk{to half of the
DDR\emph{n} standard of 3.9 or 7.8 $\mathrm{\mu}$s, which is supported by
standards at extended temperature ranges~\cite{jedec2020ddr5,
jedec2020lpddr5,jedec2014lpddr4, jedec2012ddr4, jedec2008ddr3}, or even to over}
an order of magnitude less often~\cite{venkatesan2006retention, liu2012raidr,
nair2013archshield, katayama1999fault}.

Second, the system must mitigate any errors that may occur within the small
number of DRAM cells that require frequent refreshing. Doing so requires
\amg{using either (i) additional operations to mitigate data loss (e.g., issuing
extra row activations to those cells' rows~\cite{liu2012raidr}) or (ii)
supplementary} error-mitigation mechanisms within processor (e.g.,
ECC~\cite{qureshi2015avatar} and/or bit-repair
techniques~\cite{venkatesan2006retention, lin2012secret, nair2013archshield}).
Although both \amg{approaches can} introduce new performance and energy
overheads, \mpl{the benefits of reducing unnecessary refresh operations outweigh
the overheads introduced~\cite{liu2012raidr, ohsawa1998optimizing,
wang2014proactivedram, venkatesan2006retention, lin2012secret,
nair2013archshield, ghosh2007smart, qureshi2015avatar, patel2017reach,
nguyen2021zem}. For example, Liu et al.~\cite{liu2012raidr} project that DRAM
refresh overheads cause a 187.6\% increase in the energy-per access and a 63.7\%
system performance degradation for 64~Gib chips. By reducing the overall number
of DRAM refresh operations, \amg{the authors' proposal (RAIDR) mitigates} these
overheads by 49.7\% and 107.9\%, respectively.} 

\amg{Unfortunately, this second step is difficult for the average DRAM consumer
because it requires a trustworthy method for discriminating DRAM cells' data
retention characteristics. These characteristics vary with both the DRAM chip
circuit design (e.g., random cell-to-cell variations, locations of true and
anti-cells~\cite{liu2013experimental, kraft2018improving,
patel2019understanding}) and usage characteristics (e.g., operating conditions
such as voltage and temperature, workload access patterns), so identifying which
cells to refresh more often requires access to both internal knowledge of a
given DRAM chip \emph{and} how the chip's end use will impact data retention.
Prior works propose reliability testing~\cite{liu2012raidr, patel2017reach,
khan2016parbor, khan2014efficacy, lin2012secret, mathew2017using} and
monitoring~\cite{qureshi2015avatar, han2014data, choi2020reducing,
patel2021harp} techniques to work around the lack of this knowledge, which the
separation of concerns hides from DRAM consumers. Ultimately, this means that
\amh{system-memory} cooperation to improve the standardized DRAM refresh
algorithm are discouraged today.}
\subsection{Study 3: RowHammer Mitigation}
\label{newsec:case_sec}

\amg{Section~\ref{subsubsec:mot_dram_security} discusses the severity of the
RowHammer vulnerability, motivating the need for secure \amh{defenses} beyond
\amh{those} currently implemented. System-\amh{memory} cooperative mechanisms
are capable of supplementing these defenses} from outside of the DRAM chip to
provide strong security without requiring changes to DRAM chip hardware. Such a
solution is attractive for a system designer with interest in building a secure
system because the designer can \amh{guarantee correctness using} their own
methods rather than \amh{taking the word of external
parties}~\mpk{\cite{saroiu2022price, qureshi2021rethinking}}.

\subsubsection{Solutions to Securely Mitigate RowHammer}
\label{newsubsubsec:solutions_to_rowhammer}

\amg{We classify secure RowHammer mitigations into four groups based on
categorization by recent work~\cite{yaglikci2020blockhammer}.}
\begin{enumerate}
    \item \emph{Access-agnostic} mitigation hardens a DRAM chip against
    RowHammer independently of the memory access pattern. \mpk{This includes
    increasing the overall DRAM refresh rate~\cite{kim2014flipping,
    apple2015about, aichinger2015ddr} and memory-wide error correction and/or
    integrity-checking mechanisms such as strong
    ECC~\cite{qureshi2021rethinking, cojocar2019exploiting, kim2014flipping}.
    These mechanisms are algorithmically simple but can introduce significant
    system hardware, performance, and/or energy-efficiency overheads (e.g.,
    \mpm{a large number of} additional refresh operations~\cite{kim2014flipping,
    kim2020revisiting, bhati2016dram}).} 
    
    \item \mpf{\emph{Proactive} mitigations~\cite{yaglikci2020blockhammer,
    greenfield2012throttling, mutlu2018rowhammer, kim2014flipping} adjust the
    DRAM access pattern to prevent the possibility of RowHammer errors.}
    
    \item \mpf{\emph{Physically isolating} mitigations~\cite{konoth2018zebram,
    van2018guardion, brasser2017can, saileshwar2022randomized, hassan2019crow,
    di2023copy, tatar2018defeating, brasser2016can, saxena2022aqua,
    loughlin2023siloz} physically separate data such that accesses to one
    portion of the data cannot cause RowHammer errors in another.}
    
    \item \mpf{\emph{Reactive} mitigations~\cite{kim2014flipping,
    aweke2016anvil, son2017making, seyedzadeh2018cbt, you2019mrloc,
    lee2019twice, park2020graphene, kim2014architectural, kang2020cattwo,
    bains2015row, jedec2020ddr5, bains2016distributed, bains2016row,
    devaux2021method, yaglikci2021security, marazzi2022protrr,
    kim2015architectural, marazzi2023rega, qureshi2022hydra} identify symptoms
    of an ongoing RowHammer attack (e.g., excessive row activations) and issue
    additional row \mpl{activation or} refresh operations to prevent bit-flips
    from occurring.}
\end{enumerate}

\noindent
\amg{Choosing a secure RowHammer defense for a given system depends on the
system's particular threat model and the overheads \mpk{(e.g., performance,
energy, hardware area, complexity)} it can tolerate. For example, if DRAM is
accessible only through processor cores (e.g., peripherals are incapable of
direct memory access), secure defenses may be possible solely through on-chip
cache management~\cite{lee2023nohammer}.}

\subsubsection{Application to Today's Commodity DRAM Chips}
\label{position:subsec:rh:necessary}

\amg{Unfortunately, implementing secure RowHammer defenses is discouraged in
today's DRAM chips because the separation of concerns hides the mechanics of how
RowHammer occurs from DRAM consumers. The defenses discussed throughout
Section~\ref{newsubsubsec:solutions_to_rowhammer} all require understanding one
or more of a chip's RowHammer error characteristics, which are summarized in
Table~\ref{tab:rh_mitigation_info}.} The first is known as
\hcfirst{}~\mpk{\cite{kim2020revisiting, orosa2021deeper} or RowHammer
Threshold~\cite{kim2014flipping, yaglikci2020blockhammer,
bennett2021panopticon}}, which describes the worst-case number of RowHammer
memory accesses required to induce a bit-flip. The second is known as the blast
radius~\cite{kim2014flipping, kim2020revisiting}, which describes how many rows
\mpf{are affected by hammering a single row}. The third is the DRAM's internal
physical row address mapping~\mpm{\cite{kim2014flipping, kim2012case}}, which is
\amg{used} to identify the locations of victim rows.

\begin{table}[h]
    \centering
    \small
    \begin{tabular}{l|ccc}
                           & \multicolumn{3}{c}{\textbf{Required Information}} \\
        \textbf{Strategy}  & \hcfirstbold{}  &  \textbf{Blast Radius} &  \textbf{Row Mapping } \\\hline 
        Access-Agnostic       & \cmark & & \\               
        Proactive             & \cmark & \cmark & \\         
        Physically Isolating  & \cmark & \cmark & \cmark \\                  
        Reactive              & \cmark & \cmark & \cmark
    \end{tabular}
    \caption{Information needed by each of the four RowHammer-mitigation strategies.}
    \label{tab:rh_mitigation_info}
\end{table}

All three RowHammer error characteristics vary between DRAM manufacturers,
chips, \mpk{and cells} based on a combination of random process variation, a
\amh{chip's} particular circuit design (including yield-management techniques
such as post-manufacturing repair, target row refresh, and error correcting
codes), \mpk{and operating conditions such as temperature and
voltage~\cite{kim2014flipping, park2016statistical, park2016experiments,
kim2020revisiting, orosa2021deeper, yaglikci2021security, yun2018study,
lim2016active, farmani2021rhat, olgun2023experimental, lang2023blaster}.}

\amg{Without trustworthy access this information, DRAM consumers are discouraged
from adopting secure RowHammer defenses. To work around this limitation,
proposals for secure RowHammer defenses conduct extensive experimental testing
to estimate RowHammer error characteristics that are needed to design and/or
configure their proposals. Unfortunately, similar to efforts that improve DRAM
access latency and refresh timings discussed in
Sections~\ref{subsec:limitng_assumptions_dram_ref}
and~\ref{subsec:limitng_assumptions_dram_lat}, deploying these methods in
practice is impractical for most consumers. \mpm{\amg{These observations are
consistent with prior} works~\mpk{\cite{qureshi2021rethinking, saroiu2022price,
loughlin2021stop}} \amg{that discuss} the difficulty in practically determining
and relying on this information without support from DRAM manufacturers.}}
\subsection{Study 4: Improving Memory Reliability}
\label{newsec:case_rela}

\amg{Main memory reliability is a key system design concern} \amg{because memory
errors can cause data loss or system failure if mismanaged (discussed further in
Section~\ref{subsubsec:mot_rela}). System-\amh{memory} cooperation can
supplement the memory chip with additional mechanisms to improve its base
reliability beyond what producers alone can provide.\footnote{These mechanisms
\amh{commonly} fall under the umbrella of memory reliability, availability and
serviceability (RAS) features~\cite{synopsys2015whitepaper, dell2008system,
slayman2006impact}.}}

\subsubsection{Solutions to Improve Memory Reliability}
\label{newsubsubsec:coop_solutions_for_reliability}

\amg{System-\amh{memory} cooperative solutions that DRAM consumers can implement
to improve memory reliability identify and/or address memory errors before they
impact the system at large. Hardware solutions include} rank-level
ECC~\cite{kim2015bamboo, cardarilli2000development, yoon2010virtualized,
udipi2012lot, jian2013low, kim2015frugal, nair2016xed, jian2013adaptive,
chen2015ecc, chen2013e3cc, manzhosov2021muse, patil2021dve, wang2018content},
rank-level ECC scrubbing~\mpk{\cite{han2014data, qureshi2015avatar,
choi2020reducing, sharifi2017online, alameldeen2011energy, naeimi2013sttram,
awasthi2012efficient, sridharan2015memory, rahman2021utilizing,
sharifi2017online}}, and bit repair techniques~\cite{lin2012secret,
nair2013archshield, kline2020flower, longofono2021predicting,
kline2017sustainable, schechter2010use, nair2019sudoku, zhang2017dynamic,
wang2017architecting, kim2016relaxfault}. \amg{Software-based approaches
include} retiring known-bad memory pages~\cite{mcelog2021bad, nvidia2020dynamic,
venkatesan2006retention, baek2014refresh, hwang2012cosmic, meza2015revisiting},
and predicting failures~\cite{mukhanov2019workload, baseman2016improving,
giurgiu2017predicting, lan2010study, liang2006bluegene, boixaderas2020cost}.

\amg{These solutions all enable DRAM consumers to adapt unreliable DRAM chips to
systems that require reliable main memory at reasonable cost.\footnote{
\amg{Consumers with exceptional reliability needs, such as those targeting
extreme or hostile environments (e.g., military, automotive, industrial,
extraterrestrial), may take more extreme measures \mpk{(e.g., custom
components~\mpm{\cite{agrawal1994proposed, smart2021rugged, im2016im,
infineon2022radiation, lu1989advanced, banerjee1989two, mazumder1993design,
data2022rad, 3d2022ddr4}}, redundant resources~\cite{mathew2021thermoelectric,
kobayashi2017highly, patil2021dve})} to ensure that memory errors do not
compromise their systems.}} For example, HOTH~\cite{longofono2021predicting}
supplements rank-level ECC with a cache-like hardware mechanism to track faulty
memory locations, enabling the system to detect and correct one additional error
for each ECC word (i.e., extend SECDED to 2EC3ED).}

\subsubsection{Application to Today's Commodity DRAM Chips}
\label{subsec:rela_study_adapting_commodity_chips}

\amg{Unfortunately, exposing memory errors outside of the DRAM chip is at best a
gray area within the separation of concerns between DRAM producers and
consumers.} Commodity DRAM is designed to work for a wide variety of systems at
a reasonable (albeit unspecified)\footnote{\mph{Academic works speculate that
commodity DRAM targets a bit error rate (BER) within the range of
$10^{-16}-10^{-12}$~\cite{nair2013archshield, kim2016all,
longofono2021predicting, patel2017reach}, but we are \amh{not aware} of
industry-provided values.}} error rate. \mph{In general, a
\amg{consumer who is concerned about the remaining errors they may encounter}
 must design and build their own solutions (i.e., outside of the DRAM chip) to
tolerate memory errors.\footnote{\mph{Even designers who adopt custom DRAM
solutions that sacrifice the cost advantages of commodity memory (e.g.,
high-reliability DRAM~\cite{smart2021rugged, im2016im}) may supplement the DRAM
chips with additional error-mitigation mechanisms outside of the DRAM chip}.}}

\amg{However, these solutions fundamentally rely on understanding how those
errors might manifest in the first place.} \amh{Each error-mitigation mechanism}
targets a particular \emph{error model}, which defines the scope of the errors
that it is designed to mitigate. \amh{As a result, although} a given mechanism
efficiently mitigates errors within its target error model, it may fail to do so
if errors no longer fit the model. In such cases, a different error-mitigation
mechanism \mpk{(or possibly, a combination of multiple mechanisms)} may be more
suitable.

For example, a coarse-grained approach such as page
retirement~\mpk{\cite{mcelog2021bad, nvidia2020dynamic, venkatesan2006retention,
baek2014refresh, hwang2012cosmic, meza2015revisiting}} efficiently mitigates a
small number of errors at fixed bit positions. However, page retirement exhibits
significant capacity and performance overheads at high error rates or when
mitigating errors that change positions over time~\cite{meza2015revisiting,
lee2019exploiting, mcelog2021bad}. In contrast, a fine-grained hardware-based
approach such as a block error-correcting code~\mpk{\cite{moon2005error,
richardson2008modern, roth2006introduction, clark2013error, costello1982error,
costello2004ecc}} can efficiently mitigate a limited number of
randomly-distributed errors but can fail silently (and even exacerbate the
number of errors present~\mpm{\cite{alam2021lightweight, jeong2020pair,
criss2020improving, son2015cidra, patel2019understanding, patel2020bit,
patel2021enabling}}) when its correction capability is exceeded. We conclude
that it is essential for the system designer to \mpl{know} when and how errors
occur in a given memory chip in order to make an informed choice of which
error-mitigation mechanism to use in a particular system.

Unfortunately, \amh{DRAM consumers} generally do not have access to definitive
error models for commodity DRAM chips. \amh{Obtaining this information without
cooperation from producers requires extensive reliability testing (as discussed
throughout Section~\ref{newsec:dram_modeling_and_testing}), guidance from
external (possibly untrustworthy) sources, or otherwise reasoning about memory
errors at a high level (e.g., disregarding uninteresting failure modes). As a
result, the separation of concerns between DRAM producers and consumers
effectively discourages consumers from exploring the full design space for
error-mitigation mechanisms.}

\begin{figure*}[hb]
    \centering
    \includegraphics[width=\linewidth]{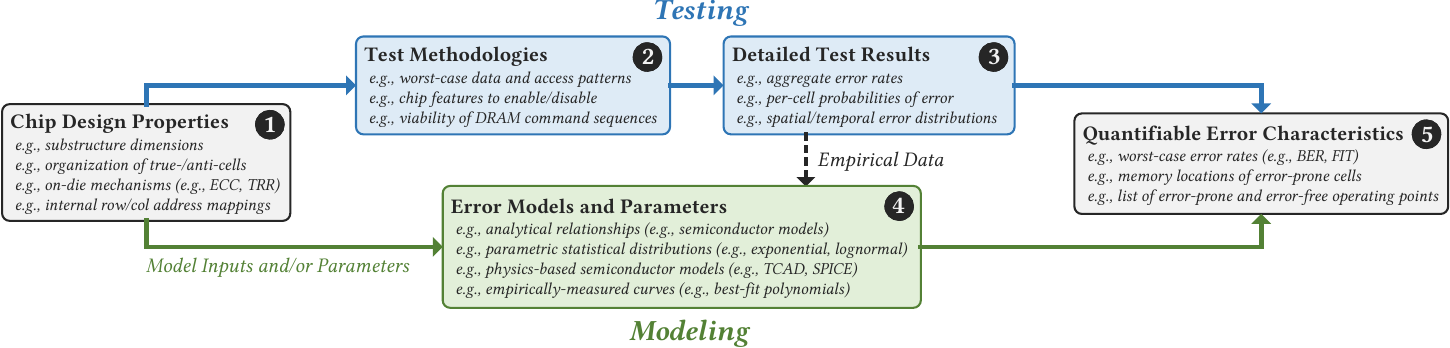}
    \caption{\mpk{Flow} of information necessary to determine key error
    characteristics for a given DRAM device.}
    \label{fig:test_flow}
\end{figure*}

\section{DRAM Reliability Testing}
\label{newsec:dram_modeling_and_testing}

\amd{As our case studies in Sections~\ref{newsec:overcoming_scaling} show,
\amg{enabling DRAM consumers to reason} about how a given DRAM operating point
way will impact its reliable operation is \emph{essential} for enabling
\amg{them to adopt system-\amh{memory} cooperative mechanisms to address DRAM
scaling challenges}. This section formalizes \mpk{the information that a DRAM
consumer may need (but does not necessarily have access to today) in order to
quantitatively reason about DRAM reliability.}}

\subsection{Information Flow During Testing}

Figure~\ref{fig:test_flow} describes the flow of information necessary for a
\amg{consumer} to quantitatively estimate\footnote{``Estimate'' because,
in general, no \mpl{model or experiment} is likely to be perfect, including
those provided by manufacturers.} a DRAM chip's error characteristics
\circled{5} starting from basic properties of the chip \circled{1}. In
principle, these characteristics can comprise \emph{any} aspect of DRAM
reliability that a \amg{consumer} wants to quantify while exploring their
system's design and/or configuration space. Examples include: (1) worst-case
error rates (e.g., bit error rate (BER) or failures in time (FIT)) across a
given set of operating points; (2) a profile of error-prone memory locations;
or (3) a list of error-free operating points \mpk{(e.g., as identified in a
shmoo analysis~\cite{baker1997shmoo})}. The error characteristics can be
estimated in two different ways: testing or modeling.

\subsubsection{\mpl{Determination from Testing}}

\mpl{First, a \amg{consumer} may estimate error characteristics using
measurements from detailed experimental testing \circled{3} across a variety of
operating conditions. Examples of measured quantities include:} aggregate error
rates, per-cell probabilities of error, and spatial/temporal error
distributions. \mpk{These measurements can be made using testing infrastructures
ranging from industry-standard large-scale testing
equipment~\cite{advantest2022t5833, teradyne2022magnum} to home-grown tools
based on commodity FPGAs~\cite{hassan2017softmc, olgun2021pidram, hou2013fpga,
kim2014flipping, gao2019computedram, chang2017understanding, ghose2018your,
weis2015retention, chang2016understanding, khan2014efficacy, wang2018dram,
ladbury2013use} or DRAM-based computing systems~\mpl{\cite{passmark2019memtest,
cojocar2020are, van2016drammer, francis2018raspberry, david2011memory}}.}

\mpl{To conduct accurate and rigorous testing, the \amg{consumer} must use an
effective test methodology \circled{2} that suits the particular DRAM chip under
test. Prior works extensively study key aspects of effective test methodologies,
including} \mpt{appropriate data and access patterns, the effects of
enabling/disabling DRAM chip features such as \mpk{target row refresh
(TRR)~\cite{frigo2020trrespass, marazzi2022protrr, hassan2021uncovering,
jattke2022blacksmith, kim2020revisiting} and on-die error correcting codes
(on-die ECC)~\mpm{\cite{nair2016xed, micron2017whitepaper, kang2014co, oh20153,
gong2017dram, son2015cidra, oh2014a, kwak2017a, kwon2014understanding,
patel2019understanding, patel2020bit, patel2021enabling}}}, and the viability of
different DRAM command sequences (e.g., sequences that enable in-DRAM row copy
operations~\cite{seshadri2013rowclone, gao2019computedram, olgun2021pidram,
chang2016low}, \mpm{true} random-number generation~\cite{olgun2021quac,
kim2019d, talukder2018exploiting, bostanci2022dr}, and physically unclonable
functions~\cite{kim2018dram, talukder2018ldpuf}).} 

\mpl{In turn, choosing an effective test methodology requires \amh{knowing}
basic} \mpt{properties about a DRAM chip's design and/or error mechanisms
\circled{1}. For example, DRAM manufacturer's design choices for the sizes of
internal storage arrays (i.e., mats~\mpm{\cite{lee2017design, zhang2014half,
son2013reducing, olgun2021quac}}), charge encoding conventions of each cell
(i.e., the true- and anti-cell organization~\cite{kraft2018improving,
liu2013experimental}), use of on-die reliability-improving mechanisms
\mpk{(e.g., on-die ECC, TRR)}, and organization of row and column addresses all
play key roles in determining \amh{whether a DRAM chip is likely to show errors
from} key error mechanisms (e.g., data retention~\cite{hamamoto1998retention,
kraft2018improving, patel2019understanding, liu2013experimental,
bacchini2014characterization, weber2005data, yamaguchi2000theoretical},
access-latency-related failures~\cite{kim2018solar, lee2013tiered,
lee2015adaptive, lee2017design, chang2016understanding, olgun2021quac,
koppula2019eden, chandrasekar2014exploiting}, and
RowHammer~\mpk{\cite{kim2014flipping, mutlu2017rowhammer, mutlu2019rowhammer,
walker2021on, yang2019trap, park2016statistical}}).}

\subsubsection{\mpl{Determination from Modeling}}

\mpl{Second, the \amg{consumer} may make predictions from analytical or
empirical error models \circled{4} based on a previous understanding of DRAM
errors (e.g., from past experiments or scientific studies).} Examples of such
error models include: analytical models based on understanding DRAM failure
modes (e.g., sources of runtime faults~\mpk{\cite{croswell2000model, das2018vrl,
cardarilli2000development, hwang2012cosmic, siddiqua2013analysis,
meza2015large})}, parametric statistical models that provide useful summary
statistics (e.g., lognormal distribution of cell data-retention
times~\cite{hamamoto1995well, hamamoto1998retention, jin2005prediction,
hiraiwa1996statistical, li2011dram, hiraiwa1998local, edri2016silicon,
kim2009new, kong2008analysis}, exponential distribution of the time-in-state of
cells susceptible to variable-retention time
(VRT)~\mpm{\cite{bacchini2014characterization, kim2015avert, qureshi2015avatar,
yaney1987meta, restle1992dram, shirley2014copula, kim2011characterization,
kim2011study, kumar2014detection, mori2005origin, ohyu2006quantitative,
khan2014efficacy, kang2014co, liu2013experimental}}), physics-based simulation models (e.g.,
TCAD~\cite{yang2019trap, synopsys2018sentaurus, duan20172d, pfaffli2018tcad,
jin2005prediction} and SPICE models~\cite{luo2020clr, hassan2019crow,
lee2017design, hassan2016chargecache, lee2013tiered, shin2019dram,
wang2018reducing, zhang2016restore, wang2020figaro}), and empirically-determined
curves that predict observations well (e.g., single-bit error
rates~\mpk{\cite{patel2017reach, qureshi2015avatar, liu2013experimental,
khan2014efficacy, khan2016parbor, park2016statistical}}). \mpl{Similar to
testing, using error models to predict error characteristics ultimately relies
on understanding the DRAM chip being tested because the accuracy of the
predictions requires choosing appropriate models and model parameters (e.g.,
through testing \circled{3} or directly from fundamental chip design properties
\circled{1}).}

\subsection{\mpl{Access to Modeling and Testing Information}}
\label{position:subsec:knowing_or_assuming}

\mpl{Figure~\ref{fig:test_flow} shows that determining a DRAM chip's error
characteristics through modeling or testing ultimately relies on understanding
the chip's fundamental design properties. This reliance can be implicit (e.g.,
inherent within a pre-existing workflow designed for a specific chip) or
explicit (e.g., chosen as part of a home-grown testing methodology). Therefore,
a \amg{consumer} must be vigilant of the information they (perhaps unknowingly)
rely upon at each step of their design process concerning commodity DRAM.}

\mpl{Fortunately, the \amg{consumer} \emph{only} needs to be concerned with the
information flow at the children of a node whose information is already known
from a trustworthy source. For example, a \amg{consumer} who wants to identify
the locations of error-prone cells (i.e., \circled{5}) using testing need not be
concerned with chip design properties (i.e., \circled{1}) if DRAM manufacturers
provide appropriate test methodologies (i.e., \circled{2}) or detailed test
results (i.e., \circled{3}).} Unfortunately, to our knowledge, neither DRAM
standards nor manufacturers provide the information in \emph{any} of the nodes
today, much less in a clear, industry-validated manner. \mpl{Therefore, the
\amg{consumer} lacks a base of trustworthy information to build upon. This
creates a barrier to entry for a \amg{consumer} who \mpm{wants} to explore
optimizations to commodity DRAM by compromising the \amg{consumer}'s ability to make
well-informed \mpm{or effective} decisions.}

In general, except for the few major DRAM customers who may be able to secure
confidentiality agreements,\footnote{Even under confidentiality, DRAM
manufacturers may be unwilling to reveal certain proprietary \ami{design
details} (e.g., on-die error correction~\mpk{\cite{patel2020bit,
gurumurthi2021hbm3}}, target row refresh~\cite{saroiu2022price}) or provide
specifically requested numbers.} \mpl{\amg{consumer}s would need to rely on
(possibly incorrect \mpm{or incomplete}) \emph{inferences} or \emph{assumptions}
based on domain knowledge or reverse-engineering studies (e.g., similar in
spirit to~\mpm{\cite{patel2017reach, kraft2018improving, liu2013experimental,
patel2020bit, kim2020revisiting, jung2016reverse, barenghi2018software,
hassan2021uncovering, farmani2021rhat, frigo2020trrespass, wang2020dramdig,
pessl2016drama, jiang2021trrscope, kim2018solar, chang2016understanding,
lee2017design, nam2023x}}) that are not verified or supported by the DRAM
industry.}\footnote{DRAM manufacturers may make assumptions during their own
testing. However, they have full transparency into their own designs (i.e, the
root node in the \mpm{information} flow), so they can make the most informed
decision.} \mpk{As a result, the \emph{need} for assumptions can discourage
practitioners from exploring the full design space even when a given design
choice is otherwise beneficial. \mpl{We conclude that} the \emph{lack of
information transparency} is a serious impediment to \amh{adopting new designs
for addressing DRAM scaling challenges today.}}
\section{Rethinking Today's Separation of Concerns}
\label{sec:spec_as_problem}

\amg{Our case studies throughout Section~\ref{newsec:overcoming_scaling}
demonstrate that although DRAM consumers have explored new and creative ways to
address DRAM scaling challenges, their solutions face serious practicality
concerns because of limited access to information about DRAM chip reliability.}
In this section, we hypothesize that the unavailability of \amg{this
information} is caused by a \emph{lack of transparency} within DRAM standards
which provide \emph{control over}, but not \emph{insight into}, DRAM operations. 

\subsection{\xmt{The Problem of Information Unavailability}}
\label{subsec:problem_of_unavailability}

\amg{Addressing DRAM scaling challenges fundamentally requires understanding how
those challenges impact system operation. Therefore, it is unsurprising that
reliability analysis and testing is central to each of the approaches we survey
in our case studies. In some cases, solutions explicitly require reliability
testing (e.g., identifying and monitoring the physical locations of error-prone
cells). Other solutions implicitly rely on the results of reliability analysis
(e.g., configuring RowHammer defenses based on a chip's degree of
vulnerability). Ultimately, deploying consumer-driven solutions to DRAM scaling
challenges requires} \amg{some degree of understanding} of how \amg{different
(representative)} operating conditions \amg{impact DRAM and overall system
reliability}.

\amg{Unfortunately, the separation of concerns does not convey this information,
which discourages consumers from adopting such solutions in practice. For
example,} recent works~\cite{qureshi2021rethinking, saroiu2022price} discuss the
pitfalls of \mpk{designing \amg{RowHammer} defense mechanisms that rely on
knowledge of how RowHammer errors behave (e.g., \hcfirst{}, dependence on a
chip's internal cell organization)}, calling into question the practicality of
accurately determining \mpk{these details given an arbitrary DRAM chip.} Knowing
or determining this information is essential to guarantee protection against
RowHammer. However, determining it \mpk{without guidance from DRAM
manufacturers} \amh{requires trusting in a home-grown testing or
reverse-engineering methodology, which ultimately relies on knowledge of
\mpk{DRAM chip details} that likely needs to be assumed or inferred (as
discussed in Sections~\ref{position:subsec:rh:necessary}
and~\ref{newsec:dram_modeling_and_testing}).}

\amg{As a result, \amh{system-memory} cooperative solutions to overcome scaling
challenges largely remain} theoretical ideas or proofs-of-concept based on
performance and reliability characteristics that are \emph{assumed},
\emph{inferred}, or \emph{reverse-engineered} from a limited set of observations
and DRAM products (e.g., using in-house experimental
studies~\mpm{\cite{lee2015adaptive, lee2017design, kim2018solar, kim2018dram,
kim2019d, talukder2018exploiting, talukder2019prelatpuf, talukder2018ldpuf,
chang2016understanding, kim2014flipping, liu2013experimental,
hassan2017softmc,gao2019computedram, olgun2021quac, chang2017understanding,
chandrasekar2014exploiting, ghose2018your, chang2017thesis, lee2016reducing,
david2011memory}}) without DRAM manufacturers' support.

\amg{Unfortunately, this lack \amh{of} a trustworthy base of information to build upon
can discourage even the most enterprising consumers from exploring new designs.
Such exploration would require weighing any potential} benefits (e.g., improved
performance, security, etc.) against both: (1) risks (e.g., failures in the
field) associated with potentially operating outside manufacturer
recommendations and (2) limitations due to compatibility with only a subset of
all commodity DRAM products (e.g., only those that have been accurately
reverse-engineered). These risks and limitations are a serious barrier to
adoption; \amg{therefore,} we conclude that the lack of information transparency
today discourages system designers from exploring alternative designs that have
been shown to provide tangible benefits.

\begin{table*}[t]
    \centering
    \small
    \setlength\tabcolsep{3pt}
    \begin{tabular}{L{5cm}|L{3.5cm}L{9cm}}
        \textbf{Design \ami{Property}} & \textbf{Reverse-Engineered By} & \textbf{Use-Case(s) Relying on Knowing the \ami{Property}} \\\hline\hline
        \begin{tabular}[c]{@{}l@{}}Cell charge encoding convention\\(i.e., true- and anti-cell layout)\end{tabular} & Testing~\cite{patel2019understanding, kraft2018improving, patel2017reach, liu2013experimental} & Data-retention error modeling and testing \mpk{for mitigating refresh overheads} (e.g., designing worst-case test patterns)~\cite{khan2017detecting, kraft2018improving, liu2013experimental} \\\hline
        On-die ECC details & Modeling and testing~\cite{patel2019understanding, patel2020bit} & Improving reliability (e.g., designing ECC within the memory controller)~\cite{son2015cidra, cha2017defect, criss2020improving, gong2018duo}, \mpk{mitigating RowHammer}~\cite{kim2020revisiting, hassan2021uncovering, cojocar2019exploiting, jattke2022blacksmith}  \\\hline
        Target row refresh (TRR) details & Testing~\cite{hassan2021uncovering, frigo2020trrespass} & Modeling and mitigating RowHammer~\cite{frigo2020trrespass, hassan2021uncovering, jattke2022blacksmith} \\\hline
        \mpk{Mapping between internal and external row addresses} & Testing~\cite{jung2016reverse, kim2020revisiting, tatar2018defeating, barenghi2018software, wang2020dramdig, lee2017design} & Mitigating RowHammer~\cite{kim2014flipping, barenghi2018software, jung2016reverse, kim2020revisiting, farmani2021rhat} \\\hline
        \mpk{Row addresses refreshed by each refresh operation} & Testing~\cite{hassan2021uncovering} & Mitigating RowHammer~\cite{hassan2021uncovering}, improving access timings~\cite{shin2014nuat, wang2018reducing} \\\hline
        Substructure organization (e.g., cell array dimensions) & Modeling~\cite{lee2017design} and testing~\cite{chang2016understanding, lee2017design, kim2018solar} & Improving DRAM access timings~\cite{lee2017design, chang2016understanding, kim2018solar} and \ami{security~\cite{loughlin2023siloz}} \\\hline
        \begin{tabular}[c]{@{}l@{}}Analytical model parameters\\(e.g., bitline capacitance)\end{tabular} & Modeling and testing~\cite{hamamoto1998retention, liu2013experimental} & \mpm{Developing and using error models for improving overall reliability~\cite{li2011dram}, mitigating refresh overheads (e.g.,~data-retention~\cite{das2018vrl, hamamoto1998retention, hiraiwa1996statistical} and VRT~\cite{restle1992dram, shirley2014copula} models), improving access timings~\cite{lee2017design}, and mitigating RowHammer~\cite{walker2021dram, park2016statistical}} \\
    \end{tabular}
    \caption{Basic DRAM chip \ami{microarchitectural design properties} that are typically assumed or inferred for experimental studies.}
    \label{tab:design_characteristics}
\end{table*}

\subsection{DRAM Standards Lack Transparency}

\amg{Historically, DRAM standards have not discussed DRAM chip reliability
because the separation of concerns assigns DRAM producers
(near)\footnote{\amg{Consumers with exceptional reliability requirements may
then choose to supplement DRAM chips with additional error-mitigation
mechanisms, as discussed in Section~\ref{subsubsec:mot_rela}.}} full
responsibility to address DRAM-related reliability concerns. Therefore, DRAM
producers are \amh{expected} to address reliability-related issues, leaving
consumers to integrate reliable DRAM chips. \amh{As technology scaling continues
to degrade DRAM chip reliability, modern DRAM standards are exposing new}
reliability-related features, such as on-die ECC scrubbing~\cite{jedec2020ddr5,
rahman2021utilizing, criss2020improving}, post-package
repair~\cite{horiguchi2011nanoscale, jedec2012ddr4, jedec2020ddr5, kim2016ecc,
wada2004post}, target row refresh~\cite{hassan2021uncovering,
frigo2020trrespass}, and refresh management~\cite{jedec2020ddr5,
saroiu2022configure}. Unfortunately, more general reasoning about reliability
remains elusive for consumers at large.}

We believe that this state of affairs naturally arises \amh{from} establishing a
a clear and explicit interface between \amg{producers and consumers}.
Consequently, ensuring that the standards leave enough flexibility for diverse
\amg{consumer} use-cases requires careful and explicit attention. This is
because the standards are susceptible to \emph{abstraction
inversion}~\cite{baker1990opening}, a design anti-pattern in which a previously
agreed-upon interface becomes an \emph{obstacle}, forcing \mpk{system designers}
to re-implement basic functionality in terms of the outdated abstraction. A
rigid interface limits what is and is not possible, potentially requiring
unproductive reverse-engineering to work around.

We \amh{contend} that \amg{the difficulty that consumers face today in
addressing DRAM scaling challenges} \mpk{clearly indicates} abstraction
inversion: \amh{the separation of concerns has} aged without sufficient
attention to flexibility. Although a fixed operating point defines a clear
interface, we believe that leaving room for (and potentially even encouraging)
different operating points \mpk{is essential today.}

\subsection{Benefits for Both Producers and Consumers}
\label{newsubsec:benefits_for_prod_and_cons}

\amg{Today's \amh{separation of concerns discourages} not only consumers from
exploring new ways to work with commodity DRAM chips but also producers from
adopting consumer-driven ideas that help address DRAM scaling challenges. In
other words, the separation of concerns effectively discourages \emph{both
parties} from exploring solutions outside their areas of concern. As a result,
neither party explores the full design space surrounding commodity DRAM chips.}

\amg{We believe that rethinking the separation to encourage cooperation stands
to benefit all aspects of DRAM technology, encompassing both the producers and
consumers who build and use DRAM, respectively. Producers gain access to a broad
base of innovation from consumers who prototype solutions (with or without
additional investment from producers themselves), \amh{thereby creating} new
opportunities for producers to make DRAM a more competitive product. Consumers
gain access to new ways to improve system-level metrics, such as performance and
energy efficiency, that were previously not practical.} \amh{Ultimately, both
producers and consumers benefit from the best possible version of DRAM
technology.}
 
\section{DRAM Standards as the Solution}
\label{sec:two_part_change_to_specs}

\amg{Separating design concerns between producers and consumers is practical for
\amh{enabling} each party to} focus on their respective areas of expertise.
However, we \amg{recommend} that the separation be crafted in a way that not
only \amg{enables both parties to help address DRAM scaling challenges}, but
ideally encourages \mpk{and aids} it. To achieve both goals, we propose
extending DRAM standards in a way that enables \amg{consumers} to make informed
decisions about how their design choices will affect \amh{a} DRAM \amh{chip's
reliable} operation. In other words, instead of modifying DRAM \emph{designs},
we advocate modifying \emph{standards} to facilitate transparency of DRAM
reliability characteristics. Armed with this information, \amg{consumers} can
freely explore how to best use commodity DRAM chips to solve their own design
challenges while preserving the separation of concerns that allows DRAM
designers to focus on building the best possible standards-compliant DRAM chips.

\subsection{\mpl{Choosing Information to Release}}
\label{position:subsec:what_to_release}

\mpg{We identify what information to release \amg{based on} our analysis of
information flow \amg{throughout DRAM reliability testing} in
Section~\ref{newsec:dram_modeling_and_testing}. We observe that, given the
information at \emph{any} node in Figure~\ref{fig:test_flow}, \amg{consumers}
can \amg{self-determine} the information at each of its child nodes. As a
result, \mpi{access to trustworthy} information at \emph{any} node \mpi{provides
\amg{consumers} with a foundation \amg{on which to build their own designs}.}
Therefore, we recommend that the DRAM industry release information at \emph{at
least one} node, \amg{but that producers be free to choose that information
based on their interests and capabilities}. This} section examines realistic
possibilities for communicating information at each \mpg{node} of the flowchart.

\subsubsection{\xmt{\ami{Microarchitectural Design Properties}}}
\label{position:subsubsec:design_characteristics}

At the lowest level, DRAM \amg{producers can} \ami{communicate} basic
\ami{microarchitectural design properties} that \amg{enable consumers} to
develop \ami{robust} test methodologies and error models. This is the most
general and flexible approach because it places no limitations on what types of
studies \amg{consumers} may pursue (e.g., in contrast to providing information
\mpk{that is useful for reasoning about} \ami{select error mechanism(s)}).
Table~\ref{tab:design_characteristics} \amg{reviews example} \ami{properties}
\amg{used by prior works to build system-level solutions for addressing DRAM
scaling challenges.} \mpk{For each \ami{design property}, we list prior works
that reverse-engineer \ami{it} and describe use-cases that rely on \ami{its}
knowledge.}

We believe that releasing \mpg{these \ami{properties} will minimally (if at
all)} impact DRAM \amg{producer}'s business interests given that each of the
\ami{properties} can be reverse-engineered with existing methods \mpk{(as shown
by Table~\ref{tab:design_characteristics}, Column 2)} and access to appropriate
tools, as demonstrated by prior studies~\mpm{\cite{patel2019understanding,
kraft2018improving, patel2017reach, liu2013experimental, patel2020bit,
hassan2021uncovering, frigo2020trrespass, jung2016reverse, kim2020revisiting,
lee2017design, chang2016understanding, kim2018solar, hamamoto1998retention,
barenghi2018software, wang2020dramdig, mukhanov2020dstress, kim2014flipping,
khan2014efficacy, farmani2021rhat}}. Releasing this information in an official
capacity confirms what is already \amg{demonstrated publicly through experiment,
yielding no further} information than \amg{others} already \amg{have the
capability to identify}. On the other hand, knowing this information empowers
\amg{\emph{all} consumers} \mpk{to confidently \amg{explore a larger design
space}, benefiting both designers and \amg{producers} in the long run (as
discussed in Section~\ref{newsubsec:benefits_for_prod_and_cons}).}

\subsubsection{Test Methodologies}

\mpk{\ami{Abstracting} beyond \ami{microarchitectural} details}, DRAM
\amg{producers} \amg{can disclose} effective test methodologies \amg{for
consumers} to \amg{conduct their own reliability studies (e.g., to explore new
viable operating points)}. Providing test methodologies absolves (1)
\amg{producers} from needing to \ami{disclose} chip design \ami{details}; and
(2) \amg{consumers} from needing the DRAM-related expertise to determine the
test methodologies from \ami{those details}.\footnote{We believe that
interested parties already have such expertise, as shown by the fact that many
studies~\mpm{\cite{patel2019understanding, kraft2018improving, patel2017reach,
liu2013experimental, patel2020bit, hassan2021uncovering, frigo2020trrespass,
jung2016reverse, kim2020revisiting, lee2017design, chang2016understanding,
kim2018solar, hamamoto1998retention, barenghi2018software, wang2020dramdig,
mukhanov2020dstress, kim2014flipping, khan2014efficacy, farmani2021rhat}}
\ami{establish} test methodologies through experimentation.}
\mpk{As a \amg{limitation}, disclosing only} test methodologies \amg{constrains}
\amg{consumers} to work with only the particular error mechanisms that the
methodologies are designed for (e.g., data-retention, RowHammer).
Table~\ref{tab:test_params} \amg{provides example test parameters} that prior
works generally \amg{depend on (e.g., assume or reverse-engineer) to conduct
reliability testing}.

\begin{table}[h]
    \centering
    \small
    \begin{tabular}{L{2.2cm}|L{5.8cm}}
        \textbf{Test Parameter} & \textbf{Description} \\\hline\hline
        Data pattern 
            & Data pattern that maximizes the chance of errors occurring~\mpk{\cite{duganapalli2016modelling, liu2013experimental, khan2014efficacy, kim2014flipping, mukhanov2020dstress, patel2017reach, kim2020revisiting, orosa2021deeper,hassan2021uncovering, cojocar2020are,tatar2018defeating, cojocar2021mfit, jattke2022blacksmith, borucki2008comparison, weis2015retention, kraft2018improving, mukhanov2020dstress,kim2018solar}} \\\hline
        Environmental conditions 
            & Temperature and voltage that lead to worst-case behavior~\mpk{\cite{park2016experiments, kim2019d, liu2013experimental, yaglikci2022understanding, orosa2021deeper, schroeder2009dram, weis2015retention,wang2018dram, yaney1987meta,hamamoto1998retention}} \\\hline
        Test algorithm 
            & Sequence of representative and/or worst-case DRAM operations to test~\mpk{\cite{cojocar2020are, kim2014flipping, liu2013experimental, lee2015adaptive, kim2018dram, kim2019d, hassan2021uncovering, jattke2022blacksmith, salman2021half}}
    \end{tabular}
    \caption{Testing parameters that are typically assumed or inferred during experimental studies.}
    \label{tab:test_params}
\end{table}

\subsubsection{Test Results and/or Error Models}
\label{position:subsubsection:test_results}

\mpk{At the highest level of abstraction}, DRAM \amg{producers} can directly
\amg{disclose} test results and/or error models related to specific studies
\amg{useful to consumers}. \amg{For example, these} could take the form of
parametric error models (e.g., the statistical relationship between operating
timings and error rates) along with parameter values for each chip;
fine-granularity error characteristics (e.g., per-column minimum viable access
timings); and/or \amg{specific} summary statistics (e.g., \hcfirst{} in studies
pertaining to RowHammer). In this way, \amg{consumers exploring new designs can
avoid the need to conduct reliability testing to identify the producer-provided
information}. \mpk{As a \amg{limitation}}, directly releasing test results
and/or error models constrains \amg{consumers} to developing solutions only for
those design concerns that pertain to the released information.
Table~\ref{tab:test_result_error_model} provides examples of key test results
and error models that prior works \mpk{leverage in order to \amg{explore new
design points based on} commodity DRAM \amg{chips}.}

\begin{table}[h]
    \centering
    \small
    \setlength\tabcolsep{3pt}
    \begin{tabular}{L{2.2cm}L{6cm}}
        \textbf{Test Result or Error Model} & \textbf{Description} \\\hline\hline
        Data-retention times
            & Minimum refresh rate required for different DRAM regions (e.g., rows, cells)~\mpk{\cite{liu2012raidr, lin2012secret, liu2013experimental,khan2014efficacy, khan2016case, kim2001block, nair2013archshield}} \\\hline
        Error profile
            & \mpk{List of cells susceptible to errors (e.g., VRT~\cite{qureshi2015avatar, liu2013experimental, khan2014efficacy}, latency-related~\cite{kim2019d, kim2018dram, kim2018solar, chang2016understanding, chandrasekar2014exploiting})} \\\hline
        Error rate summary statistics
            & \mpk{Aggregate error rates (e.g., BER~\cite{liu2013experimental, patel2017reach, patel2019understanding, weis2015retention,kang2014co}, FIT~\cite{schroeder2009dram, levy2018lessons, wang2009soft}), distribution parameters (e.g., copula~\cite{shirley2014copula}, lognormal~\cite{hamamoto1995well, hamamoto1998retention, li2011dram}, exponential~\cite{liu2012raidr,kumar2014detection})} \\\hline
        RowHammer blast radius
            & Maximum number of rows affected by hammering \mpm{one or more row(s)}~\mpk{\cite{yaglikci2020blockhammer, kim2020revisiting, yaglikci2021security, walker2021dram, loughlin2021stop, kim2014flipping}} \\\hline
        \hcfirst{} or RowHammer Threshold
            & \mpk{Minimum number of RowHammer accesses required to induce bit-flips~\cite{kim2020revisiting, orosa2021deeper, kim2014flipping, yaglikci2020blockhammer, bennett2021panopticon}}
    \end{tabular}
    \caption{\mpg{Examples of key test results and error models from prior works that study and/or optimize commodity DRAM.}}
    \label{tab:test_result_error_model}
\end{table}

\subsection{\mpl{Choosing} When to Release the Information}

\mpi{We \amg{recommend} decoupling the \emph{release} of information from the
\emph{requirement} to do so because \amg{modifying DRAM standards is a slow
process due to the need for consensus among DRAM stakeholders (discussed in
Section~\ref{newsubsec:modifying_dram_standards}).} To this end, we recommend a
practical two-step process with different approaches in the short- and
long-term.}

\subsubsection{Step 1: \mpg{Immediate Disclosure of Information}}

\mpg{We \mpi{recommend} two independent approaches to quickly release
information in the short-term. First, we recommend \amg{establishing a
publicly-accessible} database \mpi{for researchers and practitioners to
aggregate information \amg{(e.g., reverse-engineered design details) through
crowdsourcing.}} We believe this is practical given the significant
\amg{academic} and industry interest in \amg{addressing DRAM scaling
challenges}. Such a database would provide an opportunity for peer review of
posted information, increasing the likelihood that the information is
trustworthy. In the long run, we believe such a database would facilitate
information release from DRAM \amg{producers} themselves because the
\amg{producers} could simply validate database information, if not contribute
directly.}

\mpg{Second, we recommend} that commodity DRAM \amg{producers} individually
release information for current DRAM chips and those already in the field. For
example, \amg{producers} may \mpi{update} chip datasheets to incorporate
relevant design \ami{properties} or make more extensive information available
online (e.g., similar to how some \amg{producers} already provide compliance
documents and functional simulation \mpk{models} on their
websites~\mpk{\cite{micron2021dram, issi2022ddr4, nanya2022NT5AD256M16E4}}).
Releasing \mpk{the information described in
Section~\ref{position:subsec:what_to_release}} requires no changes to DRAM
designs or standards, though modifying standards (e.g., via an addendum, as we
suggest in Step 2) would help unify the information release across all
\amg{producers}. \amg{Regardless,} we believe it is important to release
information \amg{in the near term} (even if \mpk{not} standardized) so that it
is available as soon as possible.

\subsubsection{Step 2: Explicit DRAM Reliability Standards}

In the long term, we recommend \amg{modifying} DRAM standards to \mpk{promote
(or even require)} \amg{producers to disclose} information that \amg{can have
consumer-visible impact to} DRAM reliability. This may include any or all of the
information discussed throughout this \amg{paper}; we believe that the DRAM
stakeholders themselves (i.e., DRAM \amg{producers} and \amg{consumers})
\amg{must collectively} determine and standardize which information is the most
relevant and useful to regulate.

As a concrete example of how such changes to standards may occur, we reference
test methodologies~\cite{jedec2010ssdrequirements, jedec2010ssdendurance} and
error models~\cite{jedec2016failure} that JEDEC provides for NAND flash
memory endurance~\cite{cai2017error, cai2018errors, cai2012error}, including
floating-gate data retention~\cite{cai2015data, luo2018heatwatch,
luo2018improving, cai2012flash} and threshold voltage
distributions~\cite{cai2013threshold, cai2013program, cai2015read,
luo2016enabling}. These documents outline standardized best practices for
studying and characterizing endurance properties of SSD devices. We envision
analogous documents released for key DRAM error mechanisms (e.g.,
data-retention, access-timing-related, RowHammer), providing a standardized and
\amg{trustworthy} alternative to inferring the same information through
unofficial channels.

\subsection{Natural Progression Toward Transparency}

\amg{As a final note,} we anticipate that \amh{efforts to overcome} \amg{DRAM
technology scaling challenges} will naturally \amg{bring DRAM producers and
consumers closer together \amh{in pursuit of the best possible solutions}.
Diversifying consumer needs, increasing use of \amh{system-memory}
cooperation~\mpk{\cite{patterson1997case, mutlu2014research, mutlu2021primer,
kim2014flipping, mutlu2015main}}, and emerging, non-traditional DRAM
architectures~\mpm{\cite{devaux2019true, kwon202125, mutlu2021primer,
oliveira2021damov, he2020newton, niu2022184qps, ahn2016scalable, lee20221ynm,
mutlu2019processing, gomez2021benchmarking, gomez2021benchmarkingmemory}} all
challenge existing DRAM design and use practices today. \amh{As scaling
challenges continue to worsen, the opportunity costs of maintaining today's
separation of concerns will do so as well.}}

\amg{\amh{However,} if we are to step ahead of worsening DRAM scaling
challenges, we must ensure that the standards of the future \emph{proactively}
enable the whole community (both industry and academia) to \amh{collectively}
develop creative and effective solutions.} \amg{\amh{Although} recent changes to
DRAM standards such as refresh management~\cite{jedec2020ddr5,
saroiu2022configure} and on-die ECC scrubbing~\cite{jedec2020ddr5,
rahman2021utilizing, criss2020improving} are increasing signs of cooperation,
these changes are \emph{reactions} to long-standing problems. \amh{Our work
preempts this post hoc approach to system design, preemptively forging a path
toward cooperative solutions capable of holistically address the scaling
challenge.}} 
\section{Conclusion}
\label{sec:conclusion}

\amg{We show that the separation of concerns between DRAM producers and
consumers is an impediment to overcoming modern DRAM scaling challenges because
it discourages exploring the full design space around standardized designs. Our
case studies \amh{that support} this observation find that consumers' lack of
insight into DRAM reliability is the key factor \amh{discouraging} more
efficient solutions based on \amh{system-memory} cooperation. \amh{We then
analyze how consumers can obtain this insight through DRAM reliability testing,
and} we introduce a two-step approach to \amh{revise the separation of concerns
to encourage \amh{system-memory} cooperation. We start} \amg{with conceptual
changes to the separation of concerns and build toward modifying the DRAM
standards that \ami{specify} the separation. Our work is a call-to-action for more
open and flexible practices for DRAM design and use, harnessing the synergy
between researchers and practitioners to fully explore the potential of DRAM
technology.}}

\section*{Acknowledgment}
We thank the members of the SAFARI Research Group for their valuable feedback
and the constructively critical environment that they provide. We specifically
thank Geraldo F. Oliveira, Jisung Park, Haiyu Mao, Jawad Haj-Yahya, Jeremie S.
Kim, Hasan Hassan, Joel Lindegger, and Meryem Banu Cavlak for the feedback they
provided on earlier versions of this paper. We thank external experts who helped
shape our arguments, including Mattan Erez, Moinuddin Qureshi, Vilas Sridharan,
Christian Weis, and Tanj Bennett. This work was supported in part by the
generous gifts provided by our industry partners, including Google, Huawei,
Intel, Microsoft, and VMware, and support from the ETH Future Computing
Laboratory and the Semiconductor Research Corporation. A much earlier version of
this work was placed on arXiv in 2022~\cite{patel2022case}. 

\setbiblabelwidth{1000} 
\bibliographystyle{IEEEtran}
\balance
\bibliography{references}
\clearpage

\appendix



\section{DRAM Trends Survey}
\label{position:appendix_a}

We survey manufacturer-recommended DRAM operating parameters as specified in
commodity DRAM chip datasheets in order to understand how the parameters have
evolved over time. We extract values from 58 independent DRAM chip datasheets
from across 19 different DRAM manufacturers with datasheet publishing dates
between 1970 and 2021. \mpk{Appendix~\ref{position:appendix_b} lists each
datasheet and the details of the DRAM chip that it corresponds to. We openly
release our full dataset on GitHub~\cite{datasheetsurveygithub}, \mpm{which
\amh{includes} all of the raw data used in this paper, including each timing and
current parameter value, and additional fields (e.g., clock frequencies, package
pin counts, remaining IDD values) that are not presented here}.}

\subsection{\mpl{DRAM Access Timing Trends}}
\label{position:subsec:timing_trends}

We survey \amh{how the} following four DRAM timing parameters that are directly
related to DRAM chip performance \amh{evolve}.

\begin{itemize}
    \item \emph{\trcd{}}: time between issuing a row command (i.e., row activation) and a column command (e.g., read) to the row.
    \item \emph{CAS Latency (or $t_{AA}$)}: time between issuing an access to a given column address and the data being ready to access.  
    \item \emph{\tras{}}: time between issuing a row command (i.e., row activation) and a precharge command.
    \item \emph{\trc{}}: time between accessing two different rows.
\end{itemize}

\begin{figure}[H]
    \centering
    \includegraphics[width=\linewidth]{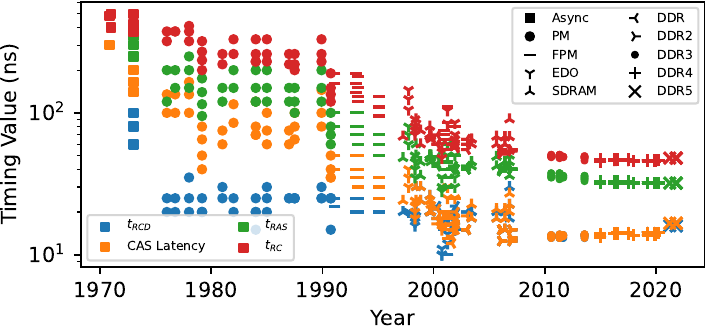}\\
    \includegraphics[width=\linewidth]{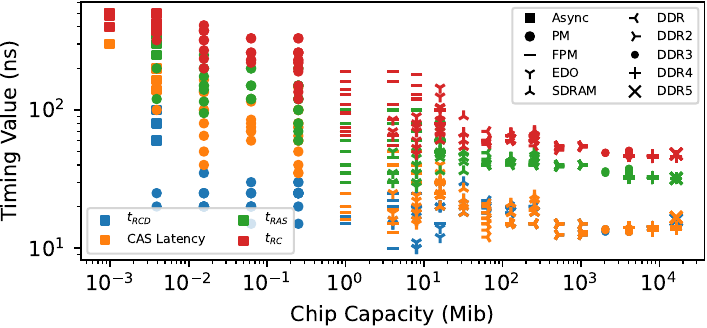}
    \caption[]{Evolution of \mpk{four} key DRAM timing parameters \mpn{(shown in
    log scale)} across years (top) and chip capacities (bottom) separated by
    DRAM standard.}
    \label{fig:da57}
\end{figure}

\begin{figure}[b]
    \centering
    \includegraphics[width=\linewidth]{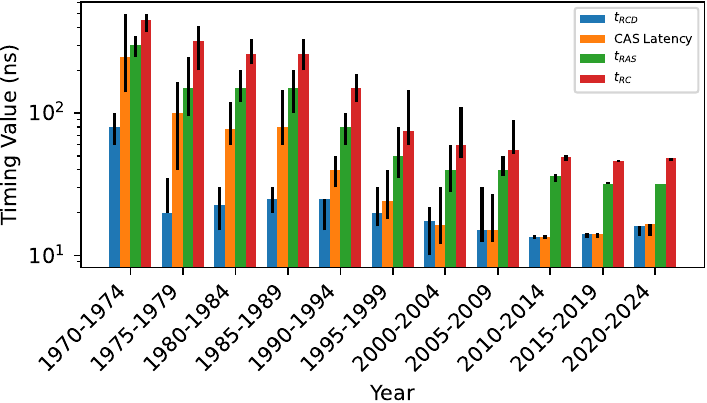}
    \includegraphics[width=\linewidth]{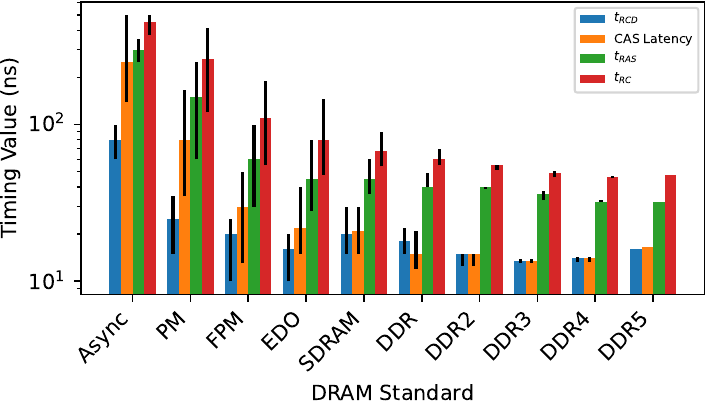}
    \caption[]{Evolution of the minimum, median, and maximum values of key DRAM timing parameters \mpn{(shown in log scale)} for each \mpk{5-year period (top) and DRAM standard (bottom)}.}
    \label{fig:da01}
\end{figure}

\noindent
Figure~\ref{fig:da57} shows how key DRAM timing parameters have evolved across
DRAM chips of different years (top) and capacities (bottom). \mpn{Timing values
are shown in log scale to better distinguish small values in newer DRAM chips.}
\mpm{Each type of marker illustrates DRAM chips of different DRAM
standards.}

We make three qualitative observations. First, \amh{although} all four DRAM
timing values roughly decrease over time, improvements \amh{are} relatively
stagnant for the last two decades (note the logarithmic Y-axis). The bulk of the
improvement in timing parameter values occurred during the period of
asynchronous DRAM, and following the introduction of SDRAM and DDR\emph{n} DRAM
chips, little to no improvements have been made despite, \mpm{or possibly as a
result of, continual} increases in overall chip storage density. Second, CAS
latency and \trcd{} converged to roughly the same values following the introduction
of synchronous DRAM. We hypothesize that this is because \mpi{similar factors
affect the latency of these operations, including a long command and data
communication latency between the external DRAM bus and the internal storage
array~\cite{keeth2007dram}.} Third, the DDR5 data points appear to worsen
relative to previous DDR\emph{n} points. However, we believe this \mpk{might be}
because DDR5 chips are new at the time of writing this article and have not yet
been fully optimized (e.g., through die revisions and other process
improvements).

To quantify the changes in \mpn{access timing \amh{values}}, we aggregate the
data points from Figure~\ref{fig:da57} by time, DRAM standard, and chip
capacity. Figure~\ref{fig:da01}, shows the minimum, median, and maximum values
\mpn{(in log scale)} for each 5-year period (top) and DRAM standard (bottom).
The data shows  that the median \trcd{}/CAS Latency/\tras{}/\trc{} reduced by
2.66/3.11/2.89/2.89\% per year on average between 1970 and 2000 but only
0.81/0.97/1.33/1.53\% between 2000 and 2015\footnote{We omit the 2020 data point
because 2020 shows a regression in CAS latency due to first-generation DDR5
chips, which we believe is not representative because of its immature
technology.} for an overall decrease of 1.83/2.10/1.99/2.00\% between 1970 and
2015.

\begin{figure*}[t]
    \centering
    \includegraphics[width=\linewidth]{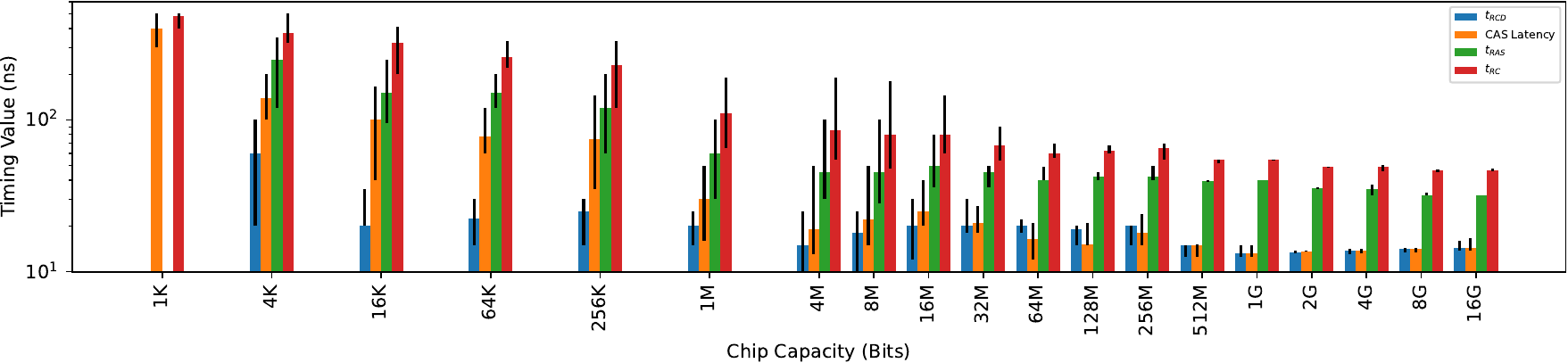}
    \caption[]{\mpm{Evolution of the minimum, median, and maximum values of key DRAM timing parameters \mpn{(shown in log scale) grouped by} DRAM chip \mpn{storage} capacity.}}
    \label{fig:da5}
\end{figure*}

\mpm{Figure~\ref{fig:da5} shows the minimum, median, and maximum of the timing
parameter values \mpn{(in log scale) grouped by DRAM chip storage
capacity.}\footnote{\mpm{We omit \trcd{} and \tras{} for the 1~Kib chips because they
do not use a row address strobe (RAS) signal.\vspace{-1ex}}} We find that the \mpn{timings
follow similar trends as in Figure~\ref{fig:da01} because higher-capacity DRAM
chips are typically introduced more recently and follow newer} DRAM standards.}

\subsection{Current Consumption Trends}

We review the evolution of the following key DRAM current consumption
measurements, which are standardized by JEDEC and are provided by manufacturers
in their datasheets.
\begin{itemize}
    \item \emph{IDD0}: current consumption with continuous row activation and
    precharge commands issued to only one bank.
    \item \emph{IDD4R}: current consumption when issuing back-to-back read operations to all banks.
    \item \emph{IDD5B}: current consumption when issuing continuous burst refresh operations.
\end{itemize}

\begin{figure}[b]
    \centering
    \includegraphics[width=\linewidth]{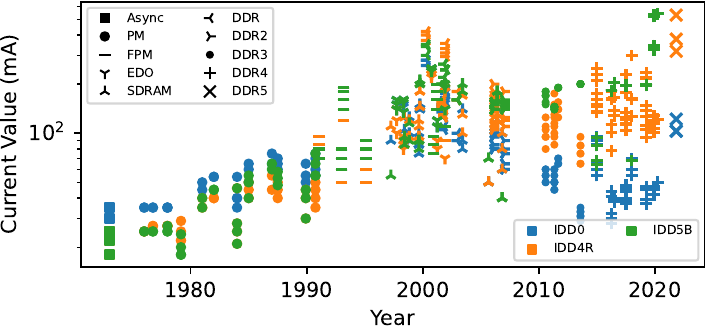}
    \includegraphics[width=\linewidth]{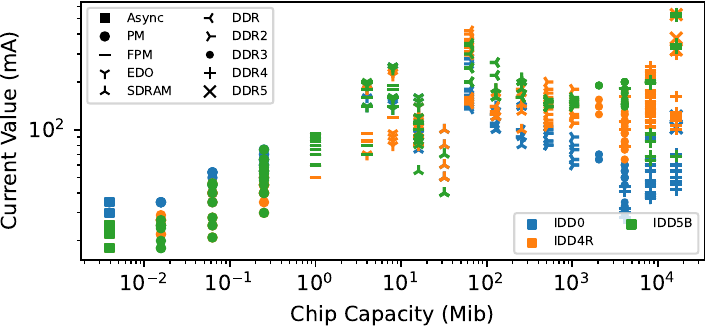}
    \caption[]{Evolution of key DRAM current consumption values (shown in log scale) across years (top) and chip capacities (bottom) separated by DRAM standard.}
    \label{fig:da46}
\end{figure}

\noindent
Figure~\ref{fig:da46} shows how key DRAM current consumption values \mpn{(in log
scale)} have evolved across DRAM chips of different years (top) and capacities
(bottom). We use different markers to show data points from chips of different
DRAM standards. \mpn{We qualitatively observe that current consumption increased
exponentially up until approximately the year 2000, which is about the time at
which improvements in access timings slowed down (as seen in
Figure~\ref{fig:da57}). After this point, different current consumption
measurements diverged as IDD0 values decreased while IDD4R and IDD5B stabilized
or increased. We explain this behavior by a change in the way DRAM chips were
refreshed as DRAM capacities continued to increase. Earlier DRAM chips refreshed
rows using individual row accesses (e.g., RAS-only refresh), which result in
comparable behavior for access and refresh operations. In contrast, newer DRAM
chips aggressively refresh \emph{multiple} rows per refresh operation (e.g.,
burst refresh), which differentiates refresh operations from normal row
accesses~\cite{mukundan2013understanding,balasubramonian2019innovations,
utah2013dram}.} 

\begin{figure}[t]
    \centering
    \includegraphics[width=\linewidth]{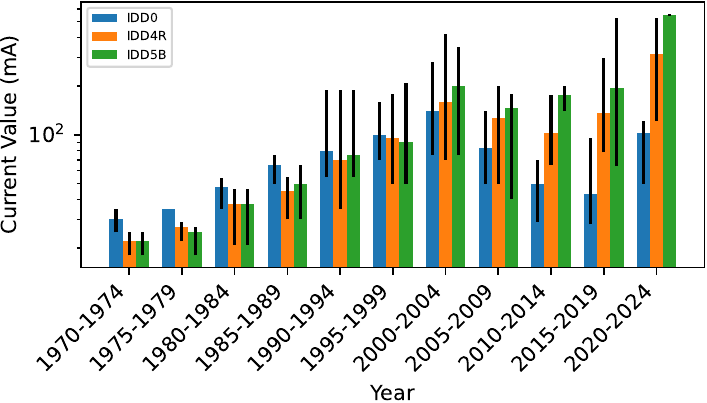}
    \includegraphics[width=\linewidth]{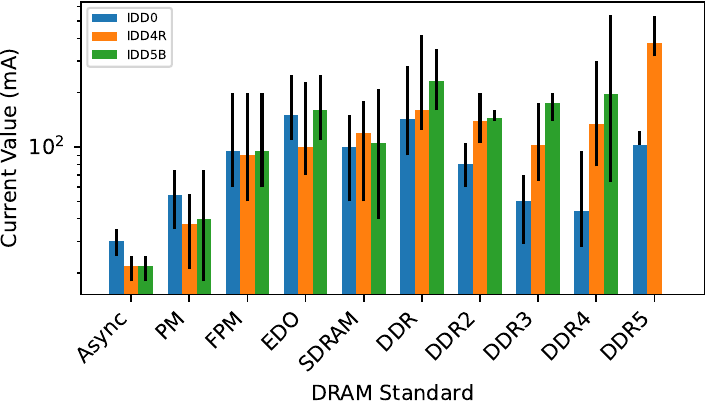}
    \caption[]{Evolution of the minimum, median, and maximum of key DRAM current consumption value \mpn{(shown in log scale)} for each \mpk{5-year period (top) and DRAM standard (bottom).\\}}
    \label{fig:da23}

    \includegraphics[width=\linewidth]{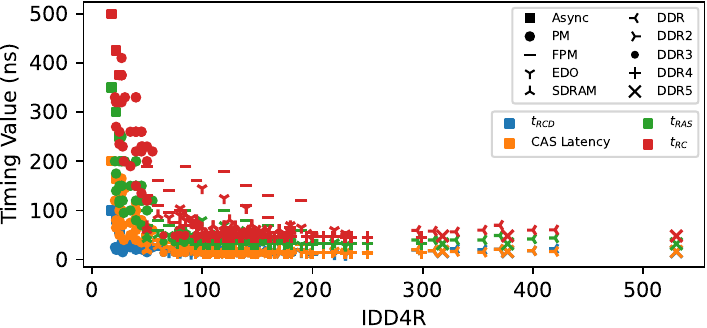}
    \captionof{figure}{\mpk{Relationship between the four timing parameters and IDD4R separated by DRAM standard.}}
    \label{fig:da9}
\end{figure}

We aggregate the current consumption data points from Figure~\ref{fig:da46} by
time and DRAM standard. Figure~\ref{fig:da23} shows the minimum, median, and
maximum values (in log scale) across each 5-year period (top) and DRAM standard
(bottom). The data shows that the median IDD0/IDD4R/IDD5B increased by
12.22/20.91/26.97\% per year on average between 1970 and 2000 but
\emph{decreased} by 4.62/1.00/0.13\% between 2000 and 2015\footnote{Similar to
Section~\ref{position:subsec:timing_trends}, we omit the 2020 data point because
the first-generation DDR5 chips exhibit outlying data values (e.g., no data
reported for IDD5B in the datasheets). for an overall increase of
0.96/11.5/17.5\% between 1970 and 2015.}

\subsection{Relationship Between Timings and Currents}

Finally, we examine the high-level relationship between the timing parameter and
current consumption values. We find that the two are generally inversely
related, which follows from the general principle that faster DRAM chips (i.e.,
lower timing parameters) require more power (i.e., increased current consumption
values). Figure~\ref{fig:da9} illustrates this relationship for the four timing
parameters studied in Section~\ref{position:subsec:timing_trends} relative to
IDD4R (i.e., the current consumption of read operations).

\subsection{\mpl{DRAM Refresh Timing Trends}}
\label{appendix_a:subsec:refresh_timings}

\mpl{DRAM refresh is governed by two key timing parameters:}

\begin{itemize}
    \item \mpl{\emph{\trefi{}} (refresh interval): time between consecutive refresh commands sent by the memory controller.}
    \item \mpl{\emph{\trfc{}}: duration of a single refresh command.}
\end{itemize}

\begin{figure}[h]
    \centering
    \includegraphics[width=\linewidth]{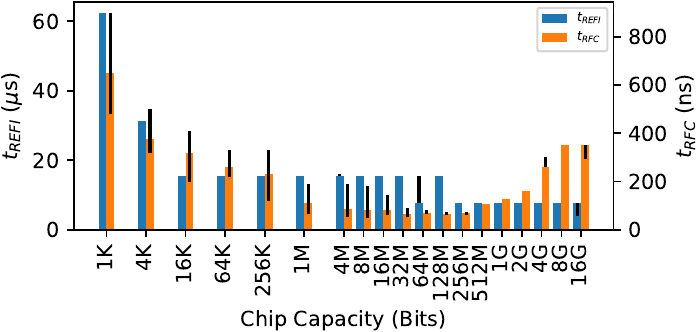}
    \caption[]{Evolution of \btrefi{} (left y-axis) and \btrfc{} (right y-axis) across DRAM chips of increasing storage capacity.}
    \label{fig:da12}
\end{figure}

\noindent
\mpl{Figure~\ref{fig:da12} shows how \trefi{} (left y-axis) and \trfc{} (right y-axis)
evolved across the DRAM chips in our study. We group chips by storage capacity
because DRAM refresh timings are closely related to capacity:
\mpm{higher-capacity} chips using the same technology require more time or more
refresh operations to fully refresh. \mpm{The error bars show the minimum and
maximum values observed across all chips for any given chip capacity.}}

\mpl{We make three observations. First, \trefi{} is shorter for
\mpm{higher-capacity} DRAM chips (e.g., 62.5~$\mu$s for an asynchronous 1~Kib
chip versus 3.9~$\mu$s for a 16~Gib DDR5 chip). This is consistent with the fact
that \mpm{higher-capacity} chips require more frequent refreshing. Second, \trfc{}
first decreases with chip capacity (e.g., 900 ns for an asynchronous 1~Kib chip
versus 54~ns for a 32~Mib SDRAM chip) but then increases (e.g., to 350~ns for a
16~Gib DDR4 chip). This is because rapid improvements in row access times (and
therefore refresh timings) initially outpaced the increase in storage capacity.
However, starting around 512~Mib chip sizes, row access times improved much more
slowly (as observed in Section~\ref{position:subsec:timing_trends}) while
storage capacity continued to increase. \amf{This matches our analysis of the
refresh penalty in Section~\ref{subsubsec:mot_dram_refresh}.} \mpm{Third, the
variation in \trfc{} across chips of each capacity (illustrated using the error
bars) decreased for \mpm{higher-capacity} chips. This is because higher-capacity
chips follow more recent DRAM standards (i.e., DDR\emph{n}), which standardize
DRAM auto refresh timings. In contrast, older DRAM chips were simply refreshed
as quickly as their rows could be accessed (e.g., every \trc{} using RAS-only
refresh).}}

\quad\\
\onecolumn

\section{\mpk{Survey Data Sources}}
\label{position:appendix_b}
Table~\ref{position:tab:sources} itemizes the 58 \mpm{DRAM} datasheets used
for our survey in Appendix~\ref{position:appendix_a}. For each datasheet, we
show the DRAM chip manufacturer, model number, DRAM standard, year, and
capacity. Our full dataset is available online~\cite{datasheetsurveygithub}.\\


\begin{minipage}[c]{\linewidth}
\begin{center}
    \small
    \renewcommand{\arraystretch}{0.93}
    \begin{tabular}{llllll}
        \textbf{Year} & \textbf{Manufacturer} & \textbf{Model Number} & \textbf{Datasheet Source} & \textbf{DRAM Standard} & \textbf{Capacity per Chip (Kib)} \\\hline
        1970 & Intel             & 1103         & \citeS{ds_intel19701103}            & Asynchronous & 1 \\
        1971 & Mostek            & MK4006       & \citeS{ds_mostek1971mk4006}         & Asynchronous & 1 \\
        1973 & Mostek            & MK4096       & \citeS{ds_mostek1973mk4096}         & Asynchronous & 4 \\
        1976 & Mostek            & MK4027       & \citeS{ds_mostek1976mk4027}         & PM           & 4 \\
        1976 & Mostek            & MK4116P      & \citeS{ds_mostek1976mk4116p}        & PM           & 16 \\
        1978 & Fairchild         & F4116        & \citeS{ds_fairchild1978f4116}       & PM           & 16 \\
        1979 & Intel             & 2118         & \citeS{ds_intel19792118}            & PM           & 16 \\
        1981 & Mitsubishi        & M5K4164ANP   & \citeS{ds_mitsubishi1981m5k4164anp} & PM           & 64 \\
        1982 & Mostek            & MK4564       & \citeS{ds_mostek1982mk4564}         & PM           & 64 \\
        1984 & NTE               & NTE4164      & \citeS{ds_nte1984nte4164}           & PM           & 64 \\
        1984 & Texas Instruments & TMS4416      & \citeS{ds_texas1984tms4416}         & PM           & 64 \\
        1985 & Mitsubishi        & M5M4256P     & \citeS{ds_mitsubishi1985m5m4256p}   & PM           & 256 \\
        1987 & Samsung           & KM41464A     & \citeS{ds_samsung1987km41464a}      & PM           & 256 \\
        1987 & Texas Instruments & TMS4464      & \citeS{ds_texas1987tms4464}         & PM           & 256 \\
        1989 & Texas Instruments & SMJ4464      & \citeS{ds_texas1989smj4464}         & PM           & 256 \\
        1990 & Intel             & 21256        & \citeS{ds_intel199021256}           & PM           & 256 \\
        1991 & Mitsubishi        & M5M44100     & \citeS{ds_mitsubishi1991m5m44100}   & FPM          & 4096 \\
        1993 & Mitsubishi        & M5M44256B    & \citeS{ds_mitsubishi1993m5m44256b}  & FPM          & 1024 \\
        1993 & Mosel Vitelic     & V404J8       & \citeS{ds_mosel1993v404j8}          & FPM          & 8192 \\
        1995 & Siemens           & HYB511000BJ  & \citeS{ds_siemens1995hyb511000bj}   & FPM          & 1024 \\
        1997 & Hyundai           & HY5118164B   & \citeS{ds_hyundai1997hy5118164b}    & EDO          & 16384 \\
        1997 & Samsung           & KM48S2020CT  & \citeS{ds_samsung1997km48s2020ct}   & SDRAM        & 16384 \\
        1998 & Micron            & MT48LC4M4A1  & \citeS{ds_micron1998mt48lc4m4a1}    & SDRAM        & 16384 \\
        1998 & Mosel Vitelic     & V53C808H     & \citeS{ds_mosel1998v53c808h}        & EDO          & 8192 \\
        1998 & Siemens           & HYB39S16400  & \citeS{ds_siemens1998hyb39s16400}   & SDRAM        & 16384 \\
        1999 & Samsung           & K4S160822D   & \citeS{ds_samsung1999k4s160822d}    & SDRAM        & 16384 \\
        1999 & Samsung           & K4S561632A   & \citeS{ds_samsung1999k4s561632a}    & SDRAM        & 262144 \\
        2000 & Amic              & A416316B     & \citeS{ds_amic2000a416316b}         & FPM          & 1024 \\
        2000 & ISSI              & IS41LV32256  & \citeS{ds_issi2000is41lv32256}      & EDO          & 8192 \\
        2000 & Samsung           & K4D623237A5  & \citeS{ds_samsung2000k4d623237a5}   & DDR          & 65536 \\
        2001 & Alliance          & AS4C256K16E0 & \citeS{ds_alliance2001as4c256k16e0} & EDO          & 4096 \\
        2001 & Alliance          & AS4C4M4FOQ   & \citeS{ds_alliance2001as4c4m4foq}   & FPM          & 16384 \\
        2001 & ISSI              & IS41C4400X   & \citeS{ds_issi2001is41c4400x}       & EDO          & 16384 \\
        2001 & Micron            & MT46V2M32    & \citeS{ds_micron2001mt46v2m32}      & DDR          & 65536 \\
        2001 & Micron            & MT46V32M4    & \citeS{ds_micron2001mt46v32m4}      & DDR          & 131072 \\
        2001 & Mosel Vitelic     & V58C265164S  & \citeS{ds_mosel2001v58c265164s}     & DDR          & 65536 \\
        2001 & TM Tech           & T224160B     & \citeS{ds_tm2001t224160b}           & FPM          & 4096 \\
        2003 & Micron            & MT46V64M4    & \citeS{ds_micron2003mt46v64m4}      & DDR          & 262144 \\
        2003 & Samsung           & K4S560432E   & \citeS{ds_samsung2003k4s560432e}    & SDRAM        & 262144 \\
        2005 & Amic              & A43L0632     & \citeS{ds_amic2005a43l0632}         & SDRAM        & 32768 \\
        2006 & Elite             & M52S32321A   & \citeS{ds_elite2006m52s32321a}      & SDRAM        & 32768 \\
        2006 & ISSI              & IS42S81600B  & \citeS{ds_issi2006is42s81600b}      & SDRAM        & 131072 \\
        2006 & Samsung           & K4T51043QC   & \citeS{ds_samsung2006k4t51043qc}    & DDR2         & 524288 \\
        2007 & Micron            & MT47H256M4   & \citeS{ds_micron2007mt47h256m4}     & DDR2         & 1048576 \\
        2010 & Samsung           & K4B4G0446A   & \citeS{ds_samsung2010k4b4g0446a}    & DDR3         & 4194304 \\
        2011 & Hynix             & H5TQ4G43MFR  & \citeS{ds_hynix2011h5tq4g43mfr}     & DDR3         & 4194304 \\
        2011 & Nanya             & NT5CB512M    & \citeS{ds_nanya2011nt5cb512m}       & DDR3         & 2097152 \\
        2013 & Samsung           & K4B4G0446A   & \citeS{ds_samsung2013k4b4g0446a}    & DDR3         & 4194304 \\
        2015 & Micron            & MT40A2G      & \citeS{ds_micron2015mt40a2g}        & DDR4         & 8388608 \\
        2016 & Hynix             & H5AN4G4NAFR  & \citeS{ds_hynix2016h5an4g4nafr}     & DDR4         & 4194304 \\
        2016 & Samsung           & K4A8G165WC   & \citeS{ds_samsung2016k4a8g165wc}    & DDR4         & 8388608 \\
        2017 & Hynix             & H5AN8G4NAFR  & \citeS{ds_hynix2017h5an8g4nafr}     & DDR4         & 8388608 \\
        2018 & Micron            & MT40A        & \citeS{ds_micron2018mt40a}          & DDR4         & 16777216 \\
        2019 & Hynix             & H5AN8G4NCJR  & \citeS{ds_hynix2019h5an8g4ncjr}     & DDR4         & 8388608 \\
        2019 & Samsung           & K4AAG045WA   & \citeS{ds_samsung2019k4aag045wa}    & DDR4         & 16777216 \\
        2020 & Samsung           & K4AAG085WA   & \citeS{ds_samsung2020k4aag085wa}    & DDR4         & 16777216 \\
        2021 & Hynix             & HMCG66MEB    & \citeS{ds_hynix2021hmcg66meb}       & DDR5         & 16777216 \\
        2021 & Micron            & MT60B1G16    & \citeS{ds_micron2021mt60b1g16}      & DDR5         & 16777216
    \end{tabular}
    \captionof{table}{List of DRAM chip datasheets used in our DRAM trends survey.}%
    \label{position:tab:sources}
\end{center}
\end{minipage}%

\clearpage
\twocolumn
\bibliographystyleS{IEEEtran}
\bibliographyS{referencesS}

\end{document}